\documentclass[aps,pra,twocolumn,showpacs]{revtex4}

\usepackage[dvips]{graphicx}
\usepackage{amsmath}
\usepackage{epsfig}
\usepackage{bbm}
\usepackage{color}

\newcommand{\ket}[1]{\left | \, #1 \right \rangle}
\newcommand{\bra}[1]{\left \langle #1 \, \right |}

\begin{document}

\title{Controlled localization of interacting bosons in a disordered optical lattice}
\author{Julio Santos, Rafael A. Molina, Juan Ortigoso, and Mirta Rodr\'iguez}
\affiliation{Instituto de Estructura de la Materia, CSIC, Serrano 121-23, 28006 Madrid, Spain}


\begin{abstract}
We show that tunneling and localization properties of interacting ultracold atoms in an optical lattice
can be controlled by adiabatically turning on a fast oscillatory force even in the presence of disorder.
Our calculations are based on the exact solution of the time-dependent Schr\"odinger equation, using the Floquet formalism. 
Implications of our findings for larger systems and the possibility of controlling the phase
diagram of disordered-interacting bosonic systems are discussed.
\end{abstract}
\maketitle
\section{Introduction}
Quantum control of many-body states is essential for new quantum
technology applications. Ultracold atoms trapped in optical lattices provide an excellent system to probe the coherent dynamical control of multiparticle  states.  A paradigm of the coherent control of cold atoms in optical lattices was provided by the superfluid (SF) to Mott-insulator (MI) transition via adiabatic ramping of the lattice potential \cite{SFMI}. More recently, a theoretical proposal \cite{Holthaus06} to drive a SF-MI transition by using a fast oscillatory force has been experimentally demonstrated \cite{Arimondo09}. In this work we analyze the role played by disorder for interacting atoms in a lattice in the presence of a fast non-resonant oscillatory potential.

The coherent dynamics of tight-binding systems in the presence of external fields is a very timely subject and includes phenomena such as Bloch oscillations, Wannier-Stark localization, caused by static fields, or dynamical localization due to the interaction with ac fields \cite{HolHone96,Korsch}. The energy eigenvectors of a quantum particle in a periodic lattice are Bloch waves. The corresponding eigenvalues form energy bands, and a single particle is delocalized over the lattice. Dunlap and Kenkre \cite{Dunlap88} showed that an external ac field localizes the particle when the ratio of the field magnitude and the field frequency is a root of the ordinary Bessel function of order zero. The use of Floquet theory gives rise to a great conceptual simplification \cite{HolHone96}; the structure of the quasienergy bands depends on the parameters of the time-dependent external perturbation and therefore by fine tuning the amplitude and frequency of the field, tunneling and the consequent localization phenomena can be manipulated. The dynamical suppression of tunneling persists in the presence of interactions \cite{Holthaus06} and it was proposed to induce localization for a single particle in the presence of disorder \cite{Holthaus95}. Here, we combine the fast oscillatory force, the interactions between atoms, and disorder for a one-dimensional tight-binding system. 

The interplay between disorder-induced localization and interactions is one of the main open issues in
condensed matter physics. In the context of bosonic atoms trapped in optical lattices it has risen much interest in recent years \cite{DeMarco09,Inguscio} and the exact ground state phase diagram of the disordered Bose-Hubbard (BH) model is still an open question \cite{Fisher89,Damski03,Burnett03,Kruger09,ReviewMaciej}. For commensurate fillings and strong interactions in the absence of disorder, there exist insulating MI states characterized by an integer fixed number of bosons per lattice site and a gap in the excitation spectrum. In the presence of disorder an additional insulating phase appears, the Bose-Glass (BG) phase characterized by a gapless excitation spectrum and finite compressibility.  For weak interactions and strong disorder an Anderson-Glass (AG) may appear where one or more condensates localize in Anderson states.

It has been shown previously that adiabatic ramping of a fast oscillatory force can drive an ordered system with commensurate filling from a superfluid regime into the localized regime as it effectively reduces and even suppresses the tunneling between neighboring sites \cite{Holthaus06}.  This force periodic in time but constant in space can be produced in the laboratory through an effective acceleration induced by a linear-in-time dephasing of the lasers forming the optical lattice \cite{Raizen98,Arimondo09}. By changing appropriately the parameters of this acceleration, its amplitude and frequency, it becomes possible to control the transitions between different states of the BH Hamiltonian. We show here that adiabatic quantum evolution towards a localized state is still possible in the presence of disorder both for commensurate and non-commensurate fillings. The structure of the quasienergy levels becomes more complex when disorder is introduced. The breakdown of degeneracies eliminates spectral gaps giving rise to a denser spectrum, and in principle adiabatic evolution could be spoiled at least for reasonable time scales. 

In section \ref{sec:model} we introduce the Hamiltonian and the Floquet theory used to describe the dynamics of the system. Section \ref{sec:results} discusses the time scales needed for inducing a transition into localized states for different ratios of the interaction and disorder strength. For commensurate filling we find that in spite of the breaking of symmetries induced by disorder, the time scale needed for adiabatic evolution is similar for an ordered and a disordered system.   Localization is not possible for an ordered system with non-commensurate filling via the adiabatic ramping of the fast oscillatory potential \cite{HolthausEPL} . We show that the breaking of symmetries induced by the disorder potential makes localization possible even at similar time scales than for the commensurate system. We also show how the adiabatic time scale shortens for increasing disorder strengths. Finally, we discuss the implications of our exact small system calculations for systems of experimentally relevant sizes. In Section \ref{sec:concl} we summarize our main conclusions.

\section{Time-dependent disordered Bose-Hubbard model \label{sec:model}}

We consider the dynamics of $N$ atoms trapped in the lowest Bloch band of an
optical lattice at sufficiently low temperatures such that it can be described by the BH model. The
corresponding Hamiltonian is (taking $\hbar=1$ throughout)
\cite{SFMI}

\begin{eqnarray}
&&{H}_0  = -J \sum_j   \left( {a}^\dagger_{j}
{a}_{j+1} +{a}^\dagger_{j+1}
{a}_{j} \right) + \frac{U}{2} \sum_j {n}_j({n}_j-1) 
\nonumber \\ && 
+\frac{W}{2} \sum_{j} \epsilon_j {n}_j + C \sum_j j^2 {n}_j,
\label{eq:BHM}
\end{eqnarray}
where ${a}_{j} $ is the bosonic destruction operator
for an atom localized in lattice site  $j=-M/2, .., M/2$  and
${n}_j={a}^{\dagger}_j{a}_j$. The parameter
$J$ is the tunneling matrix element, $U$ is the on-site repulsive interaction,  while $\epsilon_j \in [-1, 1]$ denotes a site-dependent random external potential of strength $W$ and $C$ accounts for the external parabolic trap. Depending on the ratios $W/U$ and $U/J$ the ground states of $H_0$ show very different properties and, for large enough $N$, different parameter values span the phase-diagram of the multiparticle system.

For $W=0$, one can span the SF-MI quantum phase transition by changing the ratio $U/J$. In the limit $U/J \rightarrow 0$ the tunneling dominates and the system is SF, while for $U/J \gg 1$ the repulsive interactions dominate and the system is an insulator, with a gapped spectrum. 
For $W \neq 0$ and $U/J \gg 1 $ the system is an insulator, but depending on the ratio $W/U$ and the mean occupation number $N/L$ ($L=M+1$ is the total length of the system, always odd in our calculations), the ground state is either a BG or a MI.  Both states are localized, in the sense that there is no transport through the lattice, and thus the density matrix has exponentially decaying off-diagonal terms. What distinguishes both phases \cite{Inguscio,Burnett03} is the structure of the spectrum, which is gapped for a MI while it is not for a BG. When disorder dominates $W/U>1$ the ground state is a BG, while for $W/U<1$ different theoretical methods predict a MI or a BG ground insulator state depending on the mean occupation number \cite{Fisher89,Damski03,Burnett03,Kruger09}. For small disorder, in the regime $U/J \rightarrow 0$, the system is SF with long-range coherence while in the region $U
 /W \ll 1$, the system condenses to an Anderson state \cite{Anderson} localized in the regions of strong disorder. These properties do not smear out in the presence of a shallow trapping potential.

We impose a linear potential $V(t)$ across the lattice which oscillates in time \cite{Raizen98,Arimondo09} and the total Hamiltonian 
\begin{eqnarray}
&& H(t)=H_0+V(t), \label{eq:H}\\ 
&&V(t)=V \cos (\omega t+\delta)\sum_j j {n}_j \label{eq:v}
\end{eqnarray} 
is then periodic with period $T=2\pi/\omega$. The Floquet theorem establishes that, due to the periodicity of the Hamiltonian, 
the solutions to the Schr\"odinger equation 
$i\frac{\partial}{\partial t} \ket{\Psi(t)} = H(t) \ket{\Psi(t)}$ are linear combinations of functions of the form
\begin{equation}
\ket{\Psi^e(t)}=e^{-i e t} \ket{\phi^e(t)}, \label{Psiphi}
\end{equation}
where $\ket{\phi^e(t)}=\ket{\phi^e(t+T)}$.
Inserting Eq. (\ref{Psiphi}) into the Schr\"odinger equation one obtains the
eigenvalue equation
\begin{equation}
H^{F}(t) \ket{\phi^e(t)}= e \ket{\phi^e(t)} , \label{Feigenval}
\end{equation}
where the Floquet Hamiltonian is defined as
\begin{equation}
H^{F}\equiv H(t)-i\frac{\partial}{\partial t}, \label{floquetHam1}
\end{equation}
and $e$ is the quasienergy of the system.
As pointed out by Sambe \cite{Sambe}, Eq.~(\ref{Feigenval})
can be solved using the standard
techniques developed for time-independent Hamiltonians, provided
we extend the Hilbert space to include the space of time-periodic
functions.
A suitable basis for this extended Hilbert space $\mathcal{R=H\bigotimes T}$
is $\{ \ket{\alpha} \otimes \ket{n} \}$, where
$\{\ket{\alpha} \}$ is a basis for the Hilbert space
$\mathcal{H}$ of the system, and we define $\bra{t} n \rangle=e^{in\omega
t}$ with $n$ integer, as the basis functions in the vector
space $\mathcal{T}$ of time periodic functions. In this basis, Eq.~(\ref{Feigenval})
becomes a matrix eigenvalue equation of infinite dimension with an
infinite number of eigenvalues. It can be proven that
if $e_{\alpha}$ is an eigenvalue with corresponding eigenvector
$\ket{\phi ^{e_{\alpha}}(t)}$, then $e_{\alpha} + p\omega$
with $p$ integer is also an eigenvalue with corresponding eigenvector
$\ket{\phi^{e_{\alpha} +p \omega}(t)}$= $e^{i p \omega
t}\ket{\phi^{e_{\alpha}}(t)}$. Accordingly, the eigenstates $\ket{\phi^{e_{\alpha} +p \omega}(t)}$ and $\ket{\phi^{e_{\alpha}}(t)}$ 
describe the same state in $\mathcal{H}$ apart from a global phase. Thus, in order to describe the time evolution in the Hilbert space we need only to solve the Floquet eigenvalue equation for 
$\varepsilon_\alpha \equiv e_{\alpha} \in (-\frac{1}{2}\omega , \frac{1}{2}\omega ]$ called first Brillouin zone in analogy to the theory of periodic crystals. It can be shown that there are $D$ distinct quasienergy states in the first Brillouin zone, if
the Hilbert space $\mathcal{H}$ is $D$-dimensional. The set of cyclic wave packets $\ket{\Psi^{\varepsilon_\alpha}(t)}$ in Eq. (\ref{Psiphi}) obtained from the Floquet eigenstates provides a complete orthogonal basis in $\mathcal{H}$ at each $t$. The solution of the Schr\"odinger equation is then
\begin{equation}
\ket{\Psi(t)}=U(t,t_0;V)\ket{\Psi(t_0)}, \label{eq:sol}
\end{equation}
where
\begin{equation}
U(t,t_0;V)=\sum_\alpha \ket{\Psi^{\varepsilon_\alpha}(t)} \bra{\Psi^{\varepsilon_\alpha}(t_0)}. \label{eq:u}
\end{equation}
From now on the term cyclic state refers to $\ket{\Psi^{\varepsilon_\alpha}(t_0)}$ with $t_0$ being the beginning of a period.

We are interested in a regime in which $\omega$ is smaller than the energy separation to higher energy bands, so that our single-band BH model of Eq.(\ref{eq:BHM}) is still valid, but it is larger than the rest of the relevant energy scales of our system, i.e. $\omega \gg U, J, W$. In this high-frequency regime, one can show than the effect of the time-periodic external potential in Eq. (\ref{eq:H}) is to renormalize the tunneling term.
To restore the lattice symmetry broken by the time-dependent potential it is useful to introduce a vector potential
$A(t)=\int_{t_0}^t V(t') dt'$ and transform the solutions of the time-dependent Schr\"odinger equation 
as $\tilde{\Psi}(t)=\exp{\left[-iA(t)x\right]}\Psi(t)$. Note that the tunneling terms, combination of creation and destruction operators on neighboring sites, acquire a non-trivial phase. On the other hand, the space dependent phase induced by the vector potential cancels out in the interaction and disorder terms, as they only depend on creation and
destruction operators on the same site.
In the limit $\omega \rightarrow \infty$  one can average the phase over one period and show that the time-periodic external potential in 
Eq. (\ref{eq:v}) leads for $t_0=0$ to an effective time-independent Hamiltonian similar to $H_0$ with an effective tunneling 
\cite{Dunlap88,Dittrich91,Holthaus92,Holthaus06,Creffield08}
\begin{equation}
J_{\rm eff}=J e^{\pm i V/\omega \sin \delta} \mathcal{J}_0 (V/\omega), \label{eq:Jbessel}
\end{equation}
where $\mathcal{J}_0(V/\omega)$ is the Bessel function of zero order. For finite but high-frequencies $\omega \gg U, J, W$, the Hamiltonian $H_0$ with the approximate effective tunneling rate of Eq. (\ref{eq:Jbessel}) still describes with high accuracy the cyclic states of the time periodic Hamiltonian (\ref{eq:H}) at $t_0$. For the zeroes of the Bessel function, the tunneling rate, although not exactly zero, is very much suppressed. 
Note that in this work we do exact time dependent calculations and do not use the effective Hamiltonian approximation. As shown below this approximation is very accurate and the main effect of the time-dependent potential Eq. (\ref{eq:v}) is to produce a renormalization of the effective tunneling in the lattice.

\begin{figure}
\epsfig{file=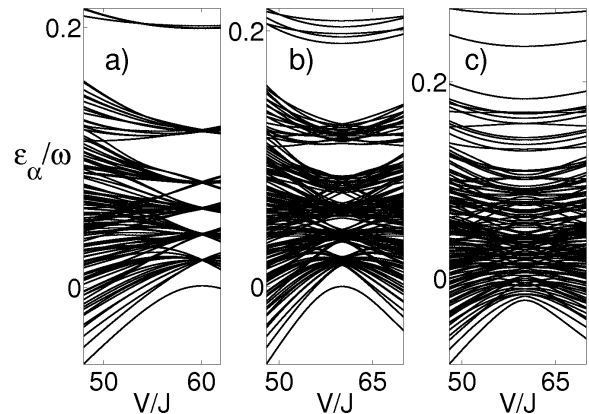,width=\columnwidth}
\caption{Quasienergy levels of the Floquet Hamiltonian as a function of the amplitude of the fast oscillatory force $V$ for different values of the disorder potential strength a) $W=0$ b) $W/J=0.1$ and c) $W/J=1$ for $N=L=5$, $U/J=0.5$, $\omega/J=25$. For high frequencies the quasienergy levels $\varepsilon_\alpha$ correspond to the eigenenergies of the static Hamiltonian $H_0$ in Eq.(\ref{eq:BHM}) with effective tunneling Eq. (\ref{eq:Jbessel}).   The structure of the quasienergy levels increases complexity for increasing values of disorder. The ground state remains gapped from the rest for all values of $W$ allowing for the adiabatic following of the ground state. The properties of the ground state change dramatically with $V$. Close to $V/\omega=2.4$ the effective tunneling is suppressed and the corresponding ground state is localized.} 
\label{fig:schemes}
\end{figure}

For an ordered lattice ($W=0$) the Hamiltonian Eq. (\ref{eq:H}) drives a transition \cite{Holthaus06, Arimondo09} from an initial SF state into a MI state by adiabatically ramping the  potential $V$ to a value $V/\omega \sim 2.4$, which corresponds to the first zero of the Bessel function, where 
$J_{\rm eff}=0$. 
In a disordered one-dimensional lattice with no interactions this oscillatory force can be used for increasing the localization of particles in the Anderson state in the regime of fast oscillations \cite{Holthaus95}, while for slow oscillations, it can be used to decrease the localization \cite{Molina06a,Molina06b}. 

Here, we combine the oscillatory driving force, the random potential, and the interatomic interactions to explore the possibility of localizing interacting particles not only in a MI state, but also in other insulating states. We explore in a controlled fashion the full parametric dependence of the ground state of interacting bosons in the presence
of disorder. As we can effectively control the value of the effective tunneling matrix elements between the lattice sites we can move between the different states in the diagonal lines of the corresponding phase diagram $U/J$ vs. $W/J$ (i.e. we keep $U/W$ constant while changing the effective $U/J$). For this control through external periodic forces to be of any use, one should 
take into account the velocity of the ramping needed for the transition between states to take place.
If one increases the amplitude of the external force adiabatically the system follows the
eigenstates (cyclic states) of the Floquet Hamiltonian. We show that adiabatic behavior can be
achieved for realistic parameters of our system. 

\begin{figure}
\centering
\begin{tabular}{cc}
\epsfig{file= 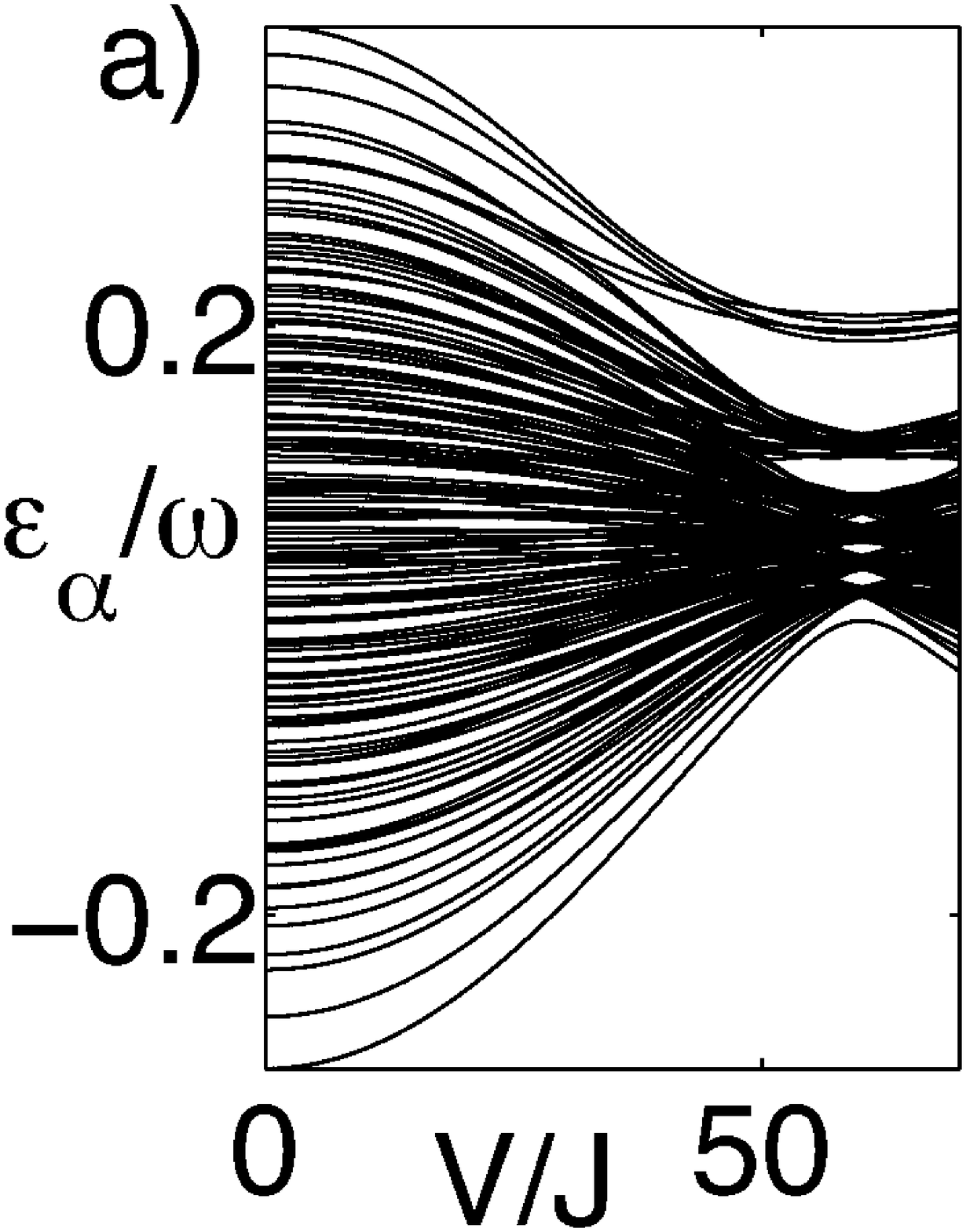,width=0.35\columnwidth} &
\epsfig{file= 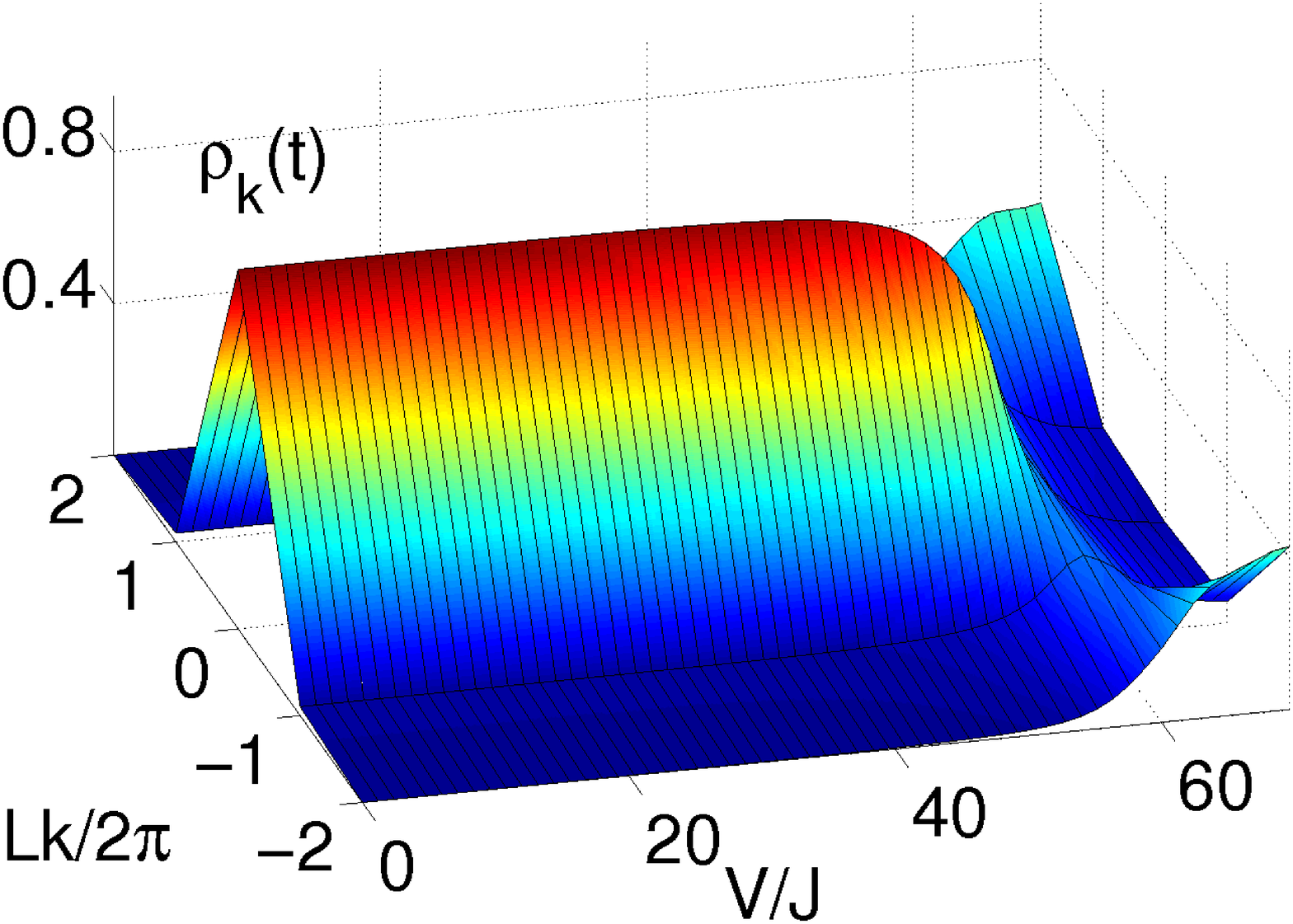,width=0.65\columnwidth} 
\end{tabular}
\caption{(Color online) a) Quasienergy levels  as a function of the amplitude of the fast oscillatory force $V$ for $N=L=5$, $U/J=0.5$, $\omega/J=25$ and $W/J=0.1$. At $V/\omega=2.4$, the effective tunneling is almost suppressed and the system localizes. The quasienergy level structure shows a gapped structure at this point. b) Evolution of the momentum distribution of the state $\ket{\Psi(t)}$ when the fast oscillatory potential is ramped linearly with step $\Delta V/J= 0.06$. The evolution is adiabatic as the state follows the ground energy eigenstate of the system. The momentum distribution changes from a peaked structure to a flat structure when the tunneling is suppressed. }
\label{fig:sfmi}
\end{figure}
\section{Adiabatic driving to localized states \label{sec:results}}

We start at $t=0$ from the ground state of $H_0$ and switch on the time-dependent potential Eq.~(\ref{eq:v}) assuming a discrete ramp $V_{m} =\Delta V m$ with $m=1,\dots, N_s$. For each $V_{m}$ we solve the Floquet eigenvalue equation and obtain the corresponding quasienergies and cyclic states. Assuming that for each $V_m$ we let the system evolve during a whole period $T$ the evolved state reads
\begin{equation}
\ket{\Psi(t=N_s T)}= \left( \prod_{m}^{N_s} U(T,0; V_m) \right) \ket{\Psi(0)},
\end{equation}
where $U$ is given by Eq.~(\ref{eq:u}). We have checked that in the adiabatic limit we obtain similar results for a continuous ramping of the external potential. According to the generalized adiabatic principle for Floquet Hamiltonians of finite size, we consider that the evolution is adiabatic when the state $\ket{\Psi(t)}$ follows the same cyclic state during the ramping
\begin{equation}
|\bra{\Psi^{\varepsilon_\alpha} (t_0,m \Delta V)} U((m-1) T, 0 ) \ket{\Psi (0)}| \sim 1
\end{equation}
for a particular $\alpha$. As we will see below, the variation of the quasienergies with potential amplitude is such that, in general, the levels become closer when $V/\omega \sim 2.4$ and the effective tunneling is coherently suppressed. In our calculations we have used only one ramping speed such that adiabatic evolution is assured in this region. Note, however, that the ramping could be faster initially for low values of $V$ and slowed down at the end. 
Calculations have been performed using the full spatial basis of $H_0$ and a truncated Fourier basis. The size of the Fourier basis is adjusted to assure convergence. We use the Arnoldi-Lanczos algorithm \cite{Lanczos} to keep only the central eigenvectors with quasienergies $\varepsilon_\alpha$. \\

To keep the problem computationally tractable we restrict ourselves to systems of size $L=5$. The results shown 
below are typical examples that demonstrate that the coherent control of 
our system is feasible. We discuss the expected behavior for increasing system sizes and conclude that adiabatic following is expected to 
occur also for realistic experimental sizes.
Note that for small system sizes, some possible phases of the disordered BH model may not appear \cite{Kruger09}. 
Also, we want to be cautious when identifying phases which occur in many-particle systems \cite{Fisher89,ReviewMaciej} with the ground states of the Hamiltonian $H_0$ for small system sizes \cite{Burnett03}. Our small system ground states are just precursor of the phases that appear in a large particle system. In this work we are interested in the time-dependent phenomena and our aim is to show that one can move adiabatically between the ground states of the effective BH Hamiltonian for varying values of the parameters $U/J$ and $W/J$ with $U/W$ fixed. This would correspond to an adiabatic transition along the diagonal lines of the phase diagram $U/J$ vs. $W/J$ for a many-particle system of larger size.

All the results shown here correspond to parameters $\delta=0$, and $C=0.002$. To illustrate the effect of disorder in the system we present 
results for one realization of the random potential $\epsilon_i=\{  0.6804,
   -0.4182,
   -0.8932,
    0.3686,
   -0.3248\}
$. Any realization of disorder that completely breaks the symmetry of the system would yield similar adiabatic time scales although
the precise value of the ramping time might depend on the exact form of the random potential.
Specifically the behavior of the time scale with occupation number and disorder ratio $W/U$ do not depend on the actual realization of the random potential.

To characterize the state $\ket{\Psi (t)}$ we use the one-particle density matrix and the momentum distribution
\begin{eqnarray}
&& \rho_{ij}=\langle \Psi| {a}_i^\dagger {a}_{j} | \Psi \rangle \label{eq:dm} \\
&& \rho_k=\sum_{i,j} e^{-i k (i-j)} \langle \Psi | {a}_i^\dagger {a}_{j} | \Psi \rangle \label{eq:mom} \\
\end{eqnarray}
where $L k/ (2 \pi)=-M/2, .., M/2$. Both quantities evolve in time, as the fast oscillatory potential is ramped up.

For an ordered system ($W=0$)  with commensurate filling, it has been shown theoretically \cite{Holthaus06} and experimentally \cite{Arimondo09} that the adiabatic ramping of the fast oscillatory potential in Eq.(\ref{eq:v}) can drive a SF-MI transition. In the presence of disorder, the reflection invariance of the system is broken, the properties of the states are different and there is no guarantee that the adiabatic time scales are within experimental reach. We show in fig. \ref{fig:schemes} the increasing complexity of the quasienergy levels with disorder strength. The breaking of symmetries makes the quasienergy spectrum more dense but strengthens the repulsion between energy levels.  
 
\begin{figure}
\centering
\begin{tabular}{cc}
\epsfig{file= 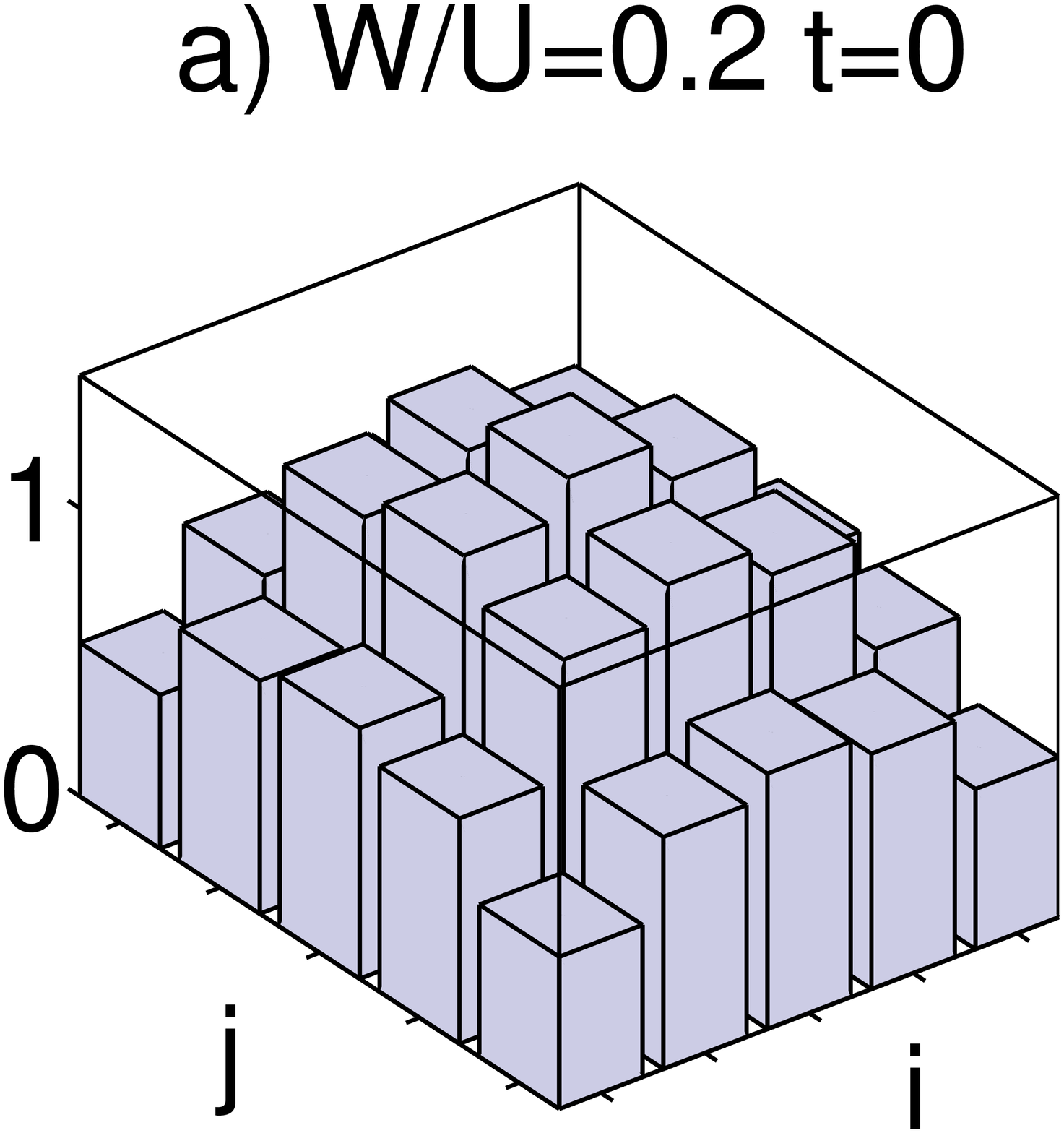,width=0.5\columnwidth} &\epsfig{file= 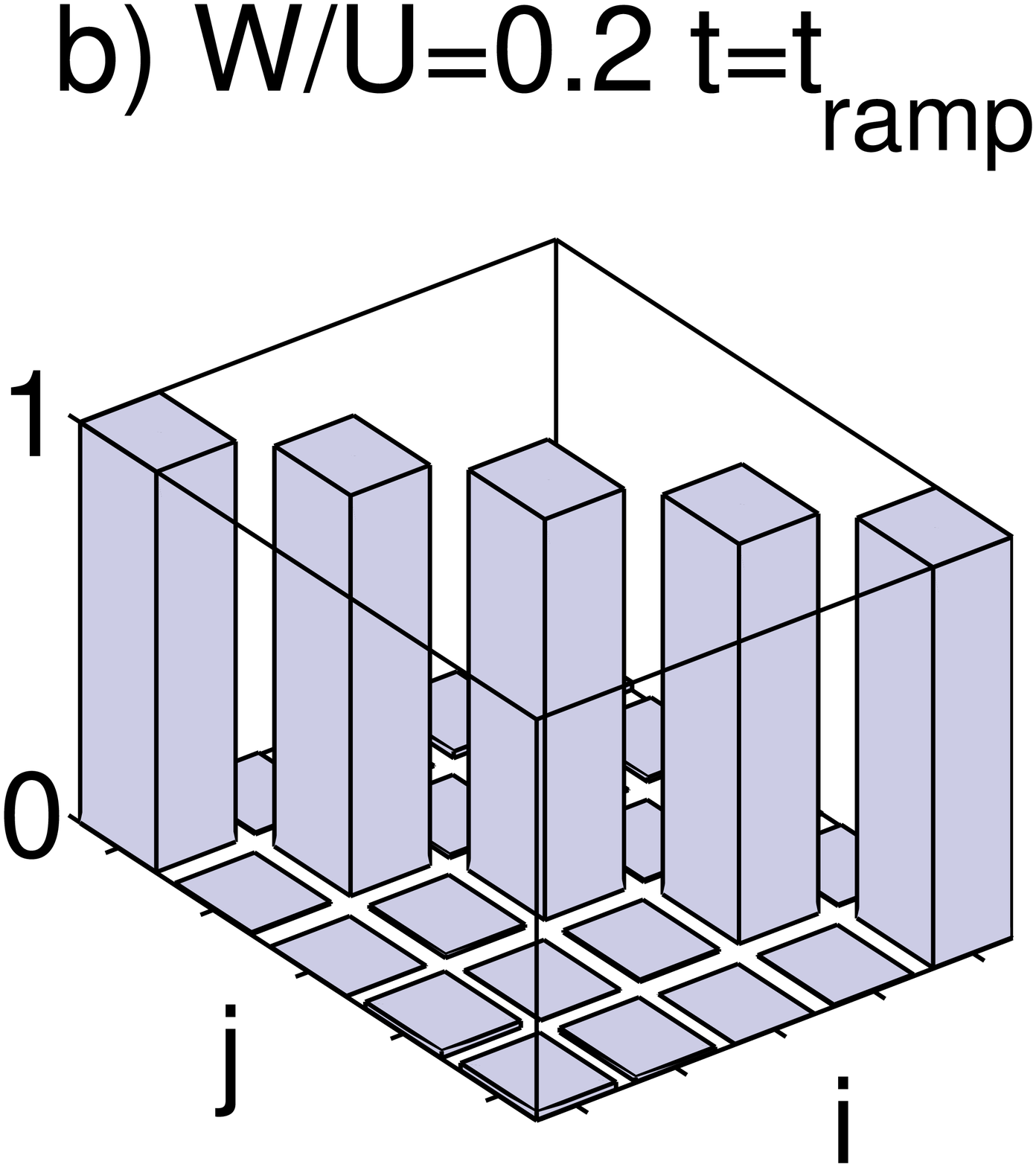,width=0.5\columnwidth} \\
\epsfig{file= 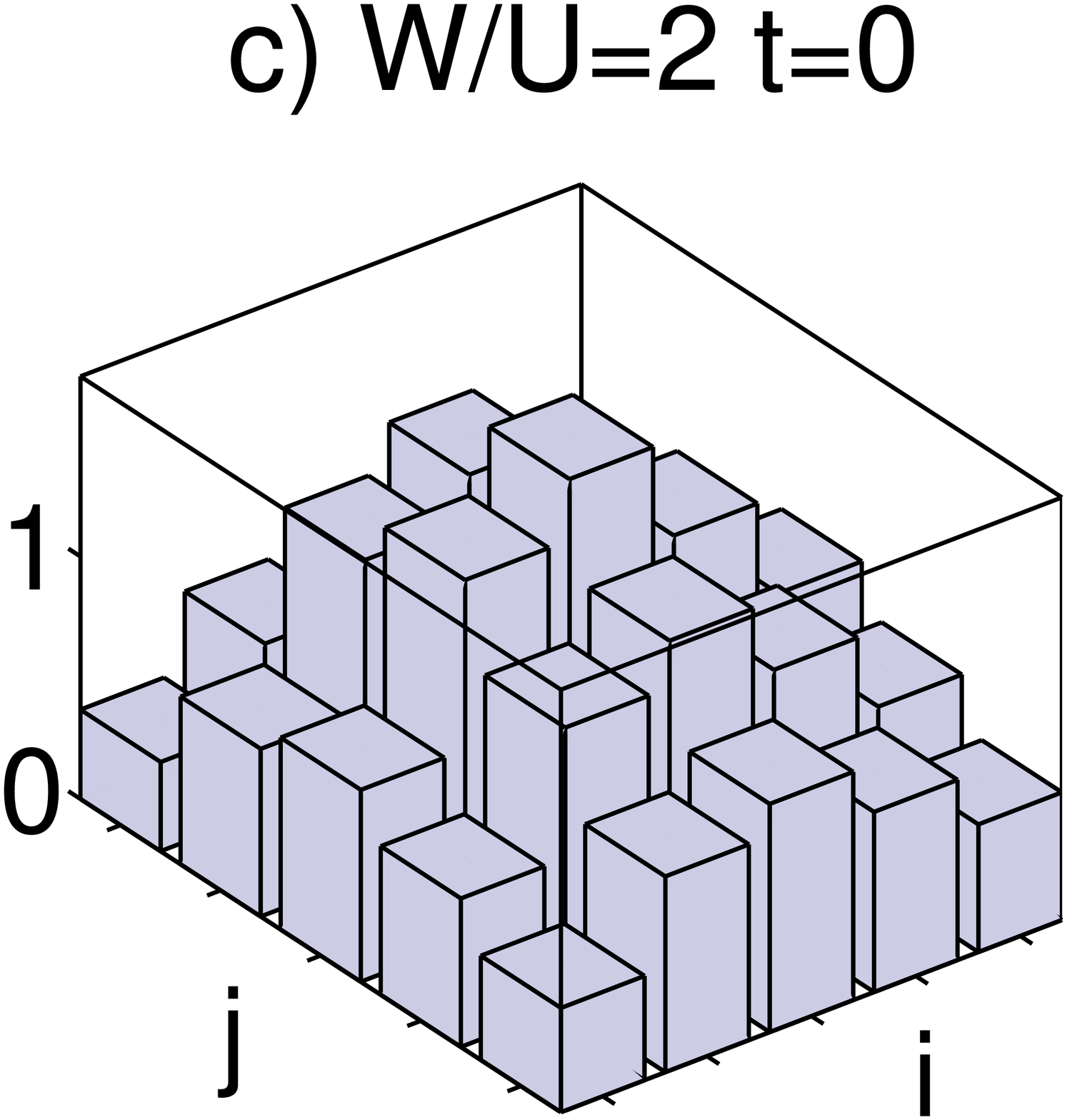,width=0.5\columnwidth} &\epsfig{file= 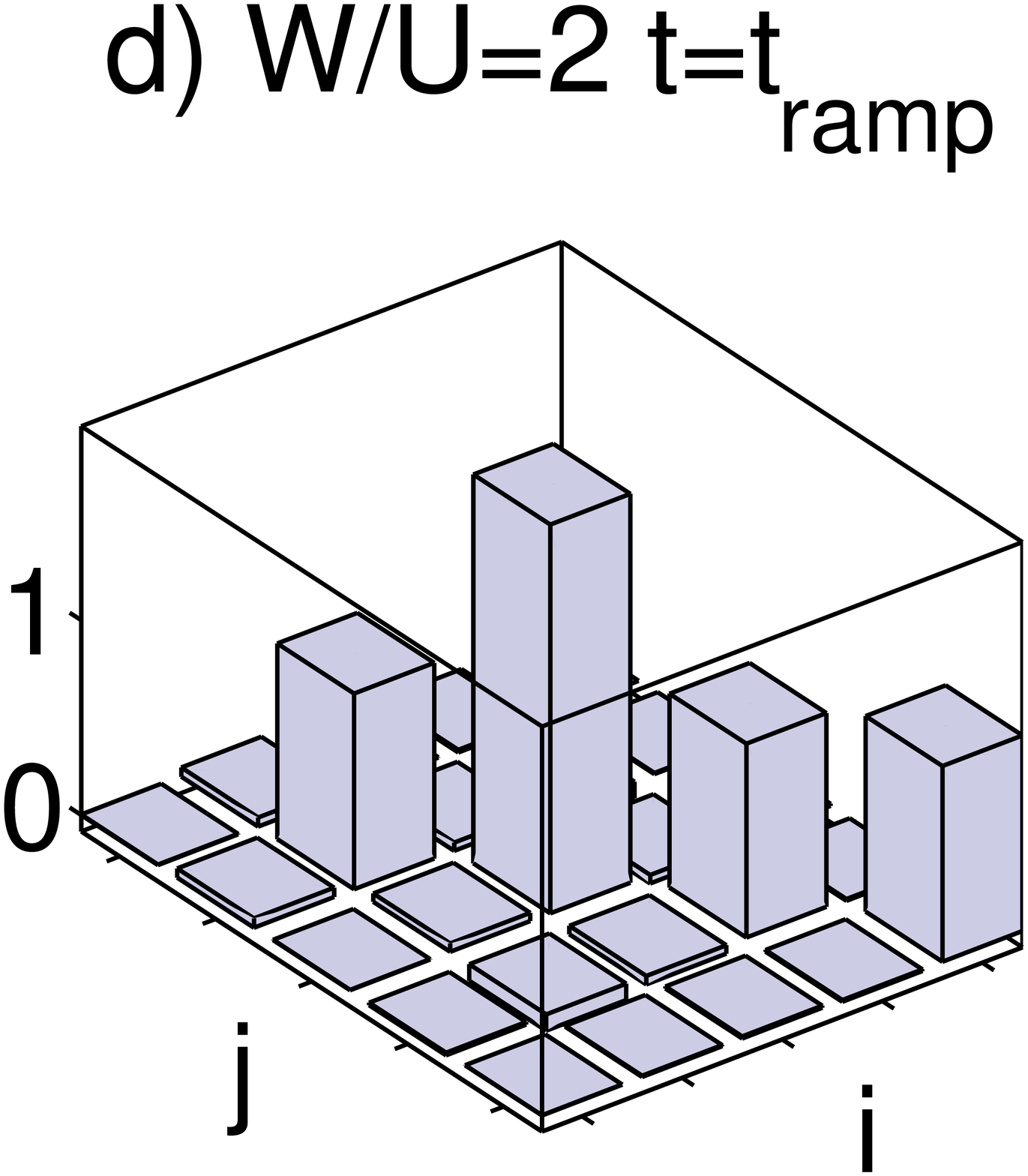,width=0.5\columnwidth} \\
\epsfig{file= 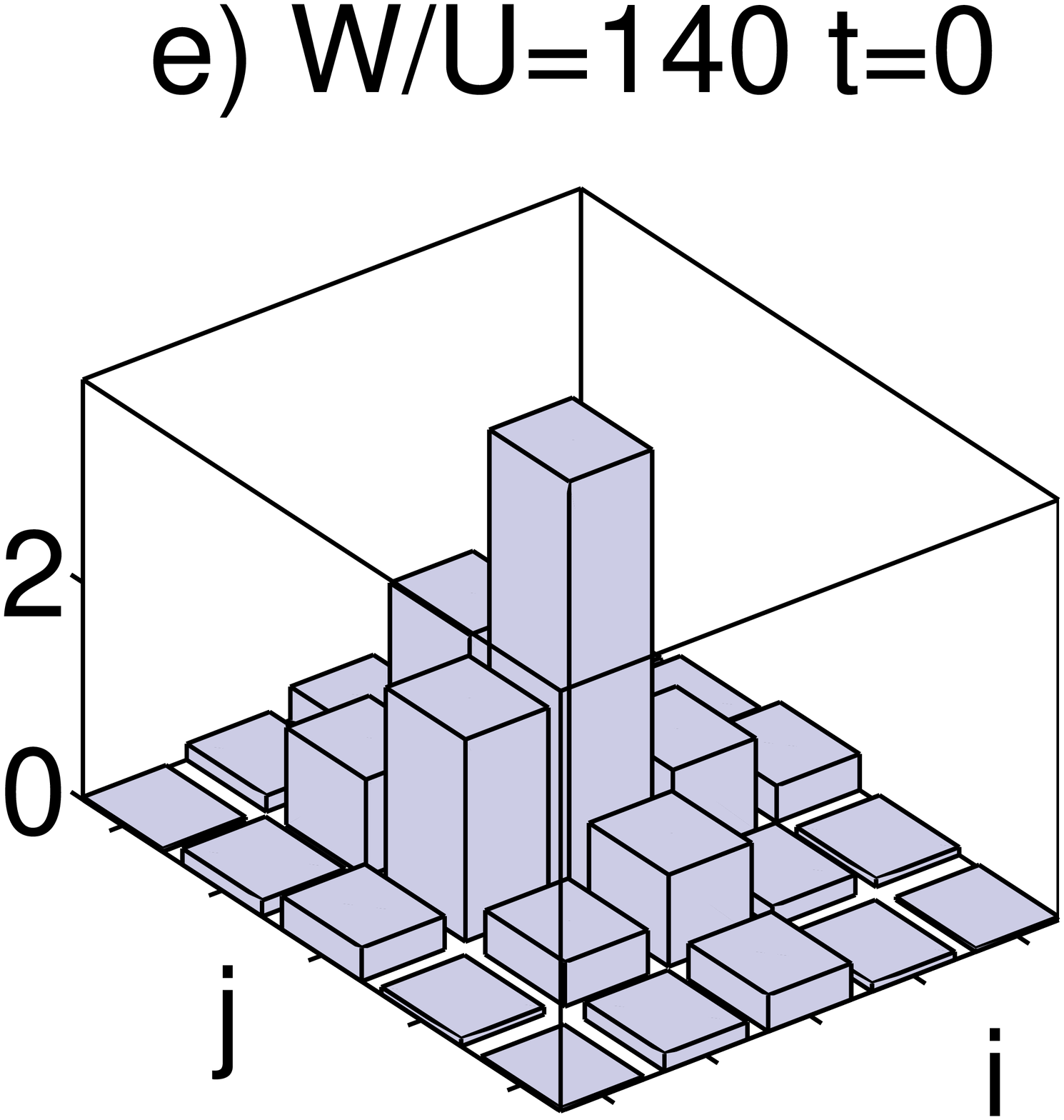,width=0.5\columnwidth,clip=} & \epsfig{file= 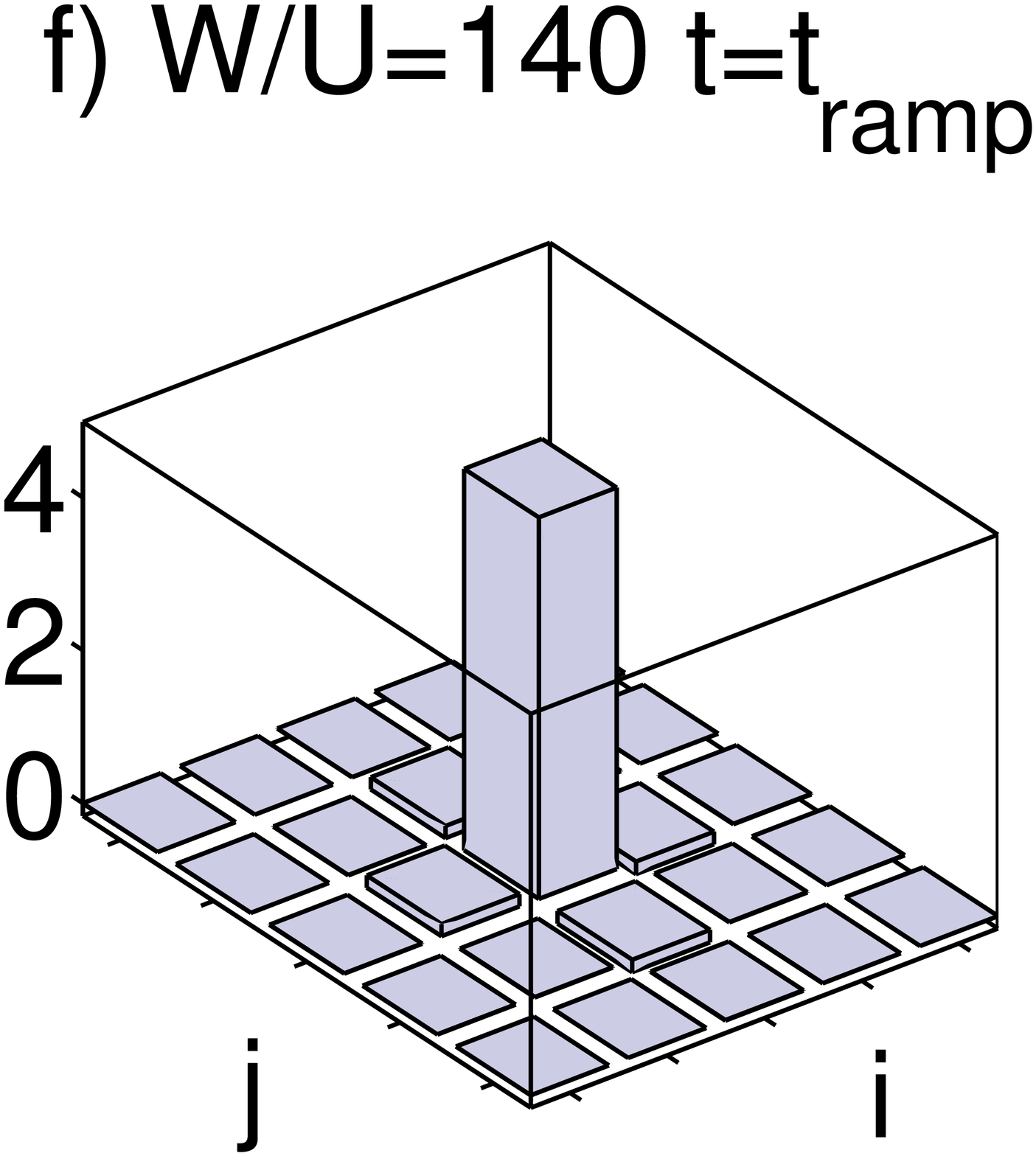,width=0.5\columnwidth} 
\end{tabular}
\caption{(Color online) One-particle density matrix $\rho_{ij}$ of the initial (left column) state when $J_{\rm eff}=J$ and final (right column) state when $J_{\rm eff}=J$ for different values of the disorder strength and $N=L=5$. a) and b) $\omega/J=25$, $W/J=0.1$, $\Delta V/J=0.06$ c) and d) $\omega/J=25$, $W/J=1$,$\Delta V/J=0.01$ e) and f) $\omega/J=37$, $W/J=7$, $\Delta V/J=0.3$.}
\label{fig:dens}
\end{figure}

\subsection{Weak disorder $W/U < 1$ \label{sec:weak}}
When the disorder is weak, in the limit $U/J\gg 0$, the ground state of the system is predicted to be a MI \cite{Burnett03,ReviewMaciej}, an insulating state with homogeneous distribution across the lattice and a gapped excitation spectrum.  Our calculations show that one can drive a transition from an initial SF state at $t=0$ into a MI-like state when we add small disorder $W/U=0.2$ to the system. The quasienergy spectrum and the momentum distribution of the evolved state are shown in figure \ref{fig:sfmi}. The one-particle density matrix smoothly changes from a flat structure to a diagonal shape for $V/\omega$ close to the first zero of the effective tunneling in Eq.~(\ref{eq:Jbessel}) as shown in fig. \ref{fig:dens} a) and b).

\begin{figure}
\begin{tabular}{cc}
\epsfig{file= 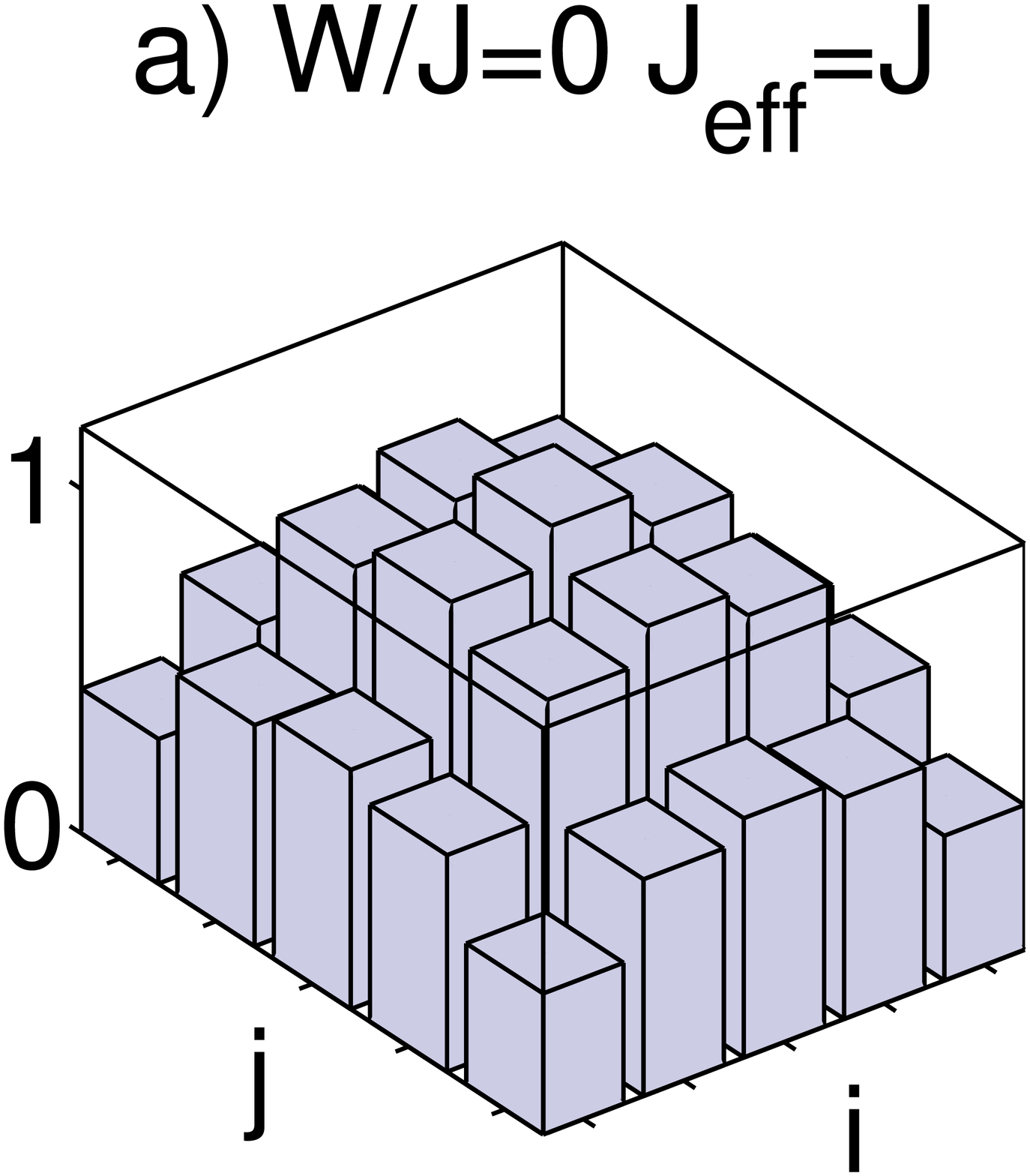,width=0.5\columnwidth} &\epsfig{file= 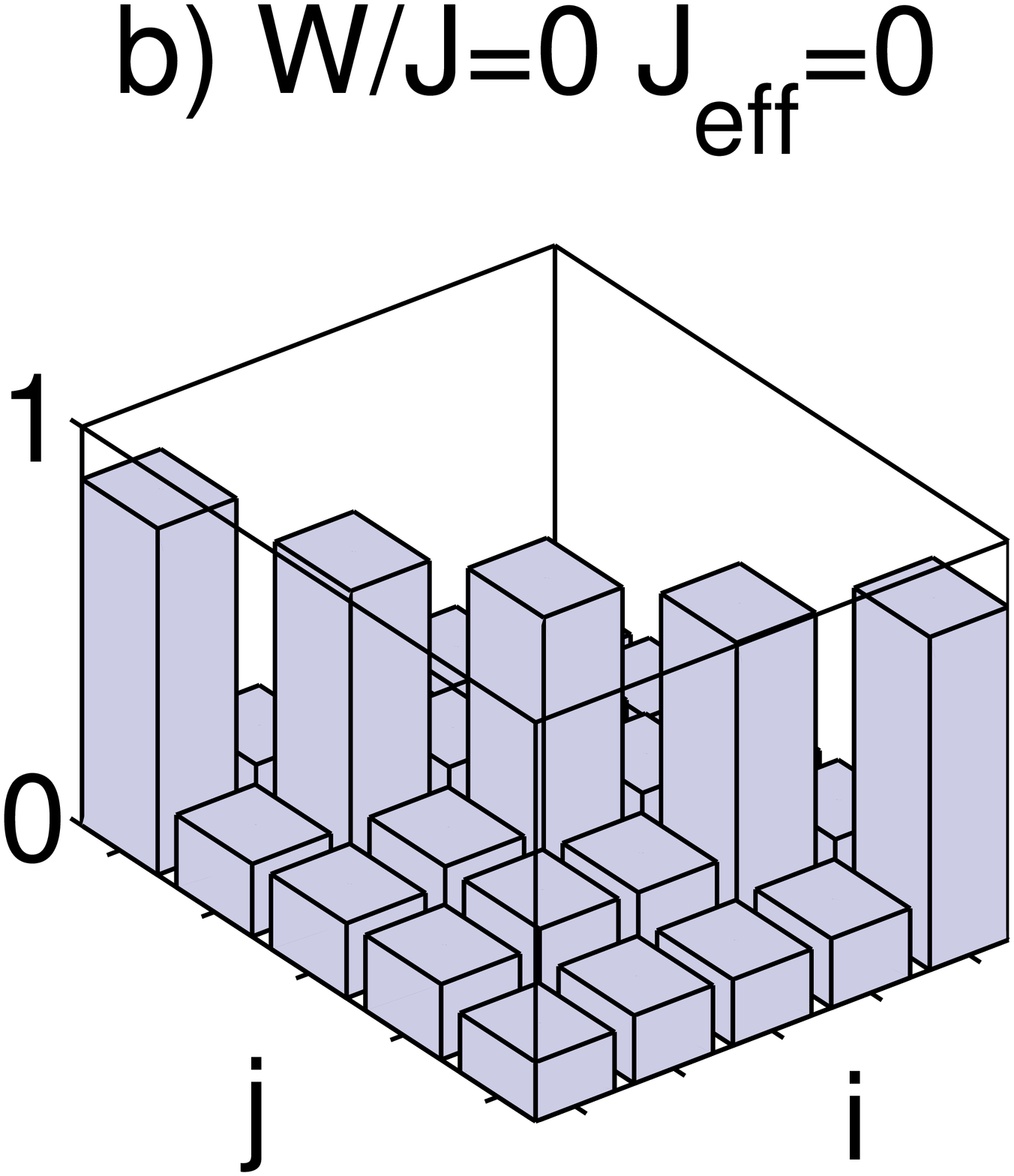,width=0.5\columnwidth} \\
\epsfig{file= 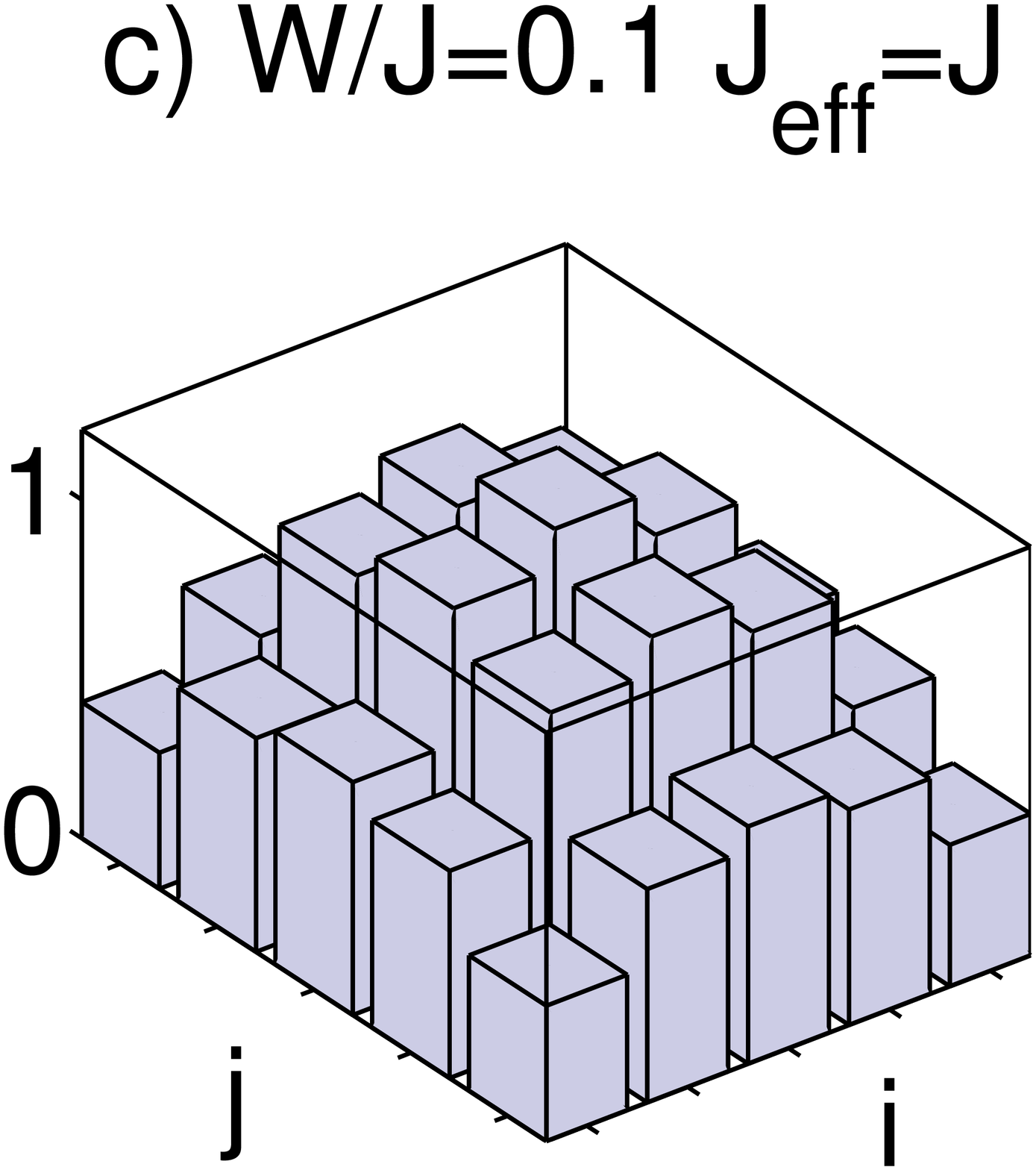,width=0.5\columnwidth} &\epsfig{file= 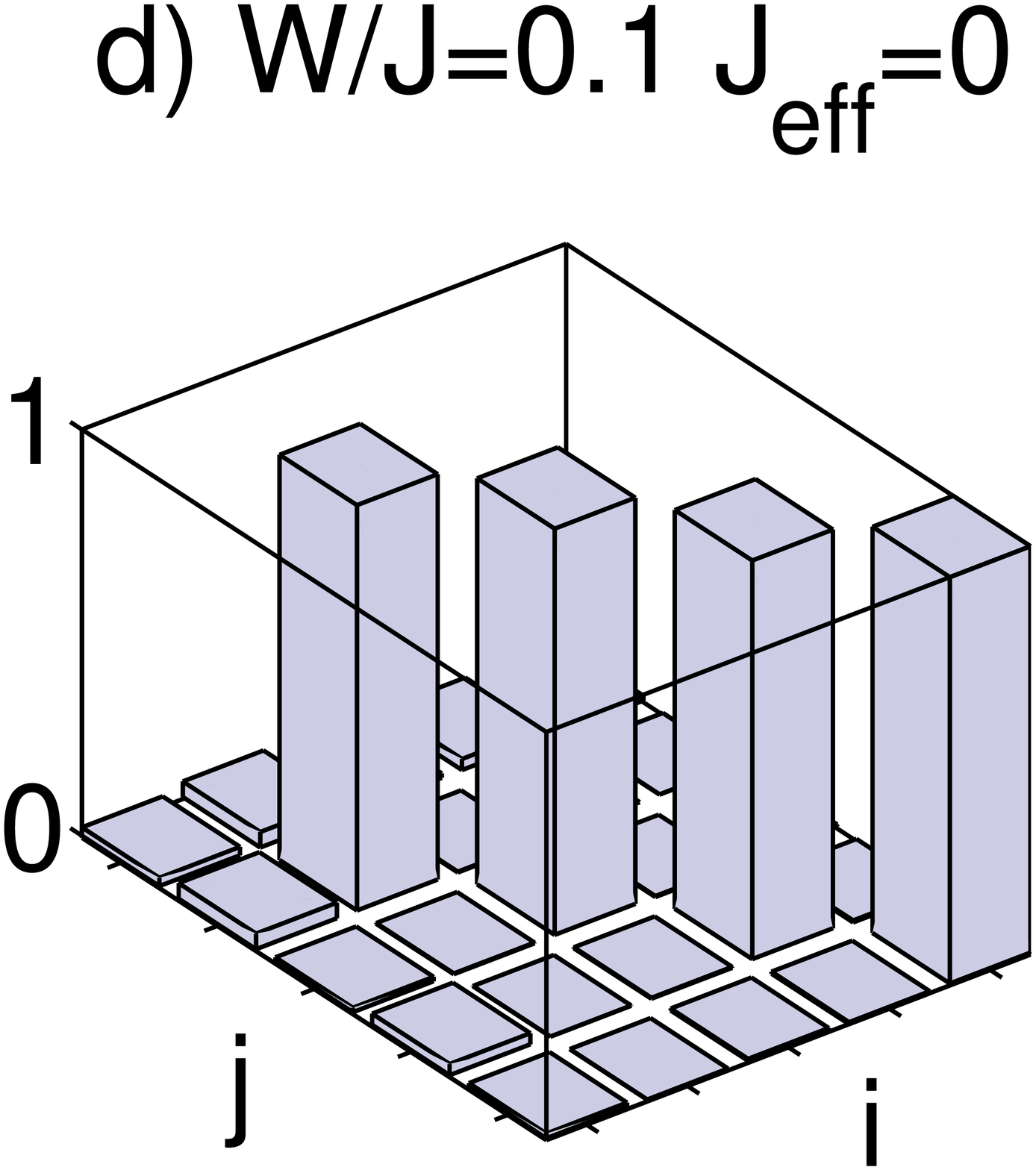,width=0.5\columnwidth} \\
\end{tabular}
\caption{(Color online) One-particle density matrix $\rho_{ij}$ of the initial (left) state at $t=0$ and final states (right) at $t=t_{\rm ramp}$ with and without disorder for $L=5$, $N=4$, $\omega/J=25$ and $U/J=0.5$ . The initial states in figures a) and c) have very similar long-range coherence properties, while only the final state in d) with finite disorder $W$ is localized.} 
\label{fig:incom}
\end{figure}

Commensurate filling implies that the ground MI state at $J_{\rm eff}=0$ is non-degenerate. As shown in fig. \ref{fig:sfmi}, the lowest quasienergy state is always gapped from the next excited states enabling the adiabatic transition. Remarkably this behavior persists in the presence of weak disorder and we obtain similar time scales for $W=0$ and for $W/U=0.2$ . Our calculations for $N=5$ yield $\Delta V/J=0.06$ which implies $N_s=2.4 \times \omega/\Delta V=6000$ and total ramping times for $V(t_{\rm ramp})= 2.4 \omega$ given by $$\Delta V t_{\rm ramp}= 2 \pi\times 2.4. $$ We thus obtain $J t_{\rm ramp} \sim 250$ which for typical $J \sim kHz$ leads to experimentally achievable ramping times of the order of tenths of seconds. 

For non-commensurate fillings localization is not possible for the proposed scheme for $W=0$ \cite{HolthausEPL}. We show in fig. \ref{fig:incom} the one-particle density matrix of the initial and final state for $N=4$ and $W=0$ and $W/U=0.2$.  Figures \ref{fig:incom} a) and c) show that both initial states have long-range coherence. The final state for $W=0$ (fig. \ref{fig:incom} b)) has long range coherence and indeed is a symmetrized combination of all the Fock states with one particle per site and one hole.  On the contrary, if we add disorder to the system, the ground state for $J_{\rm eff}=0$ is the Fock state with one particle per site and the vacancy located in the site with the highest disorder $\epsilon_i$ as shown in fig. \ref{fig:incom} d).  Due to the presence of disorder, the degeneracy is thus lifted and one achieves adiabatic passage into this localized state for a ramping with potential step 
$\Delta V/J=0.001$.

Calculations with large particle numbers \cite{Fisher89,Kruger09} predict a BG phase between the SF and the MI states for intermediate values of $J/U$ and $W/U <1$. As shown in \cite{Kruger09} these BG states disappear when the system is small. We do not observe this intermediate state in our calculations in agreement with the results in \cite{Burnett03}. For larger systems where the intermediate BG phase appears, the fast oscillatory force Eq. (\ref{eq:v}) could be used to reach either MI states or BG states by just controlling the final value of the ramping potential.

\begin{figure}
\centering
\begin{tabular}{cc}
\epsfig{file= 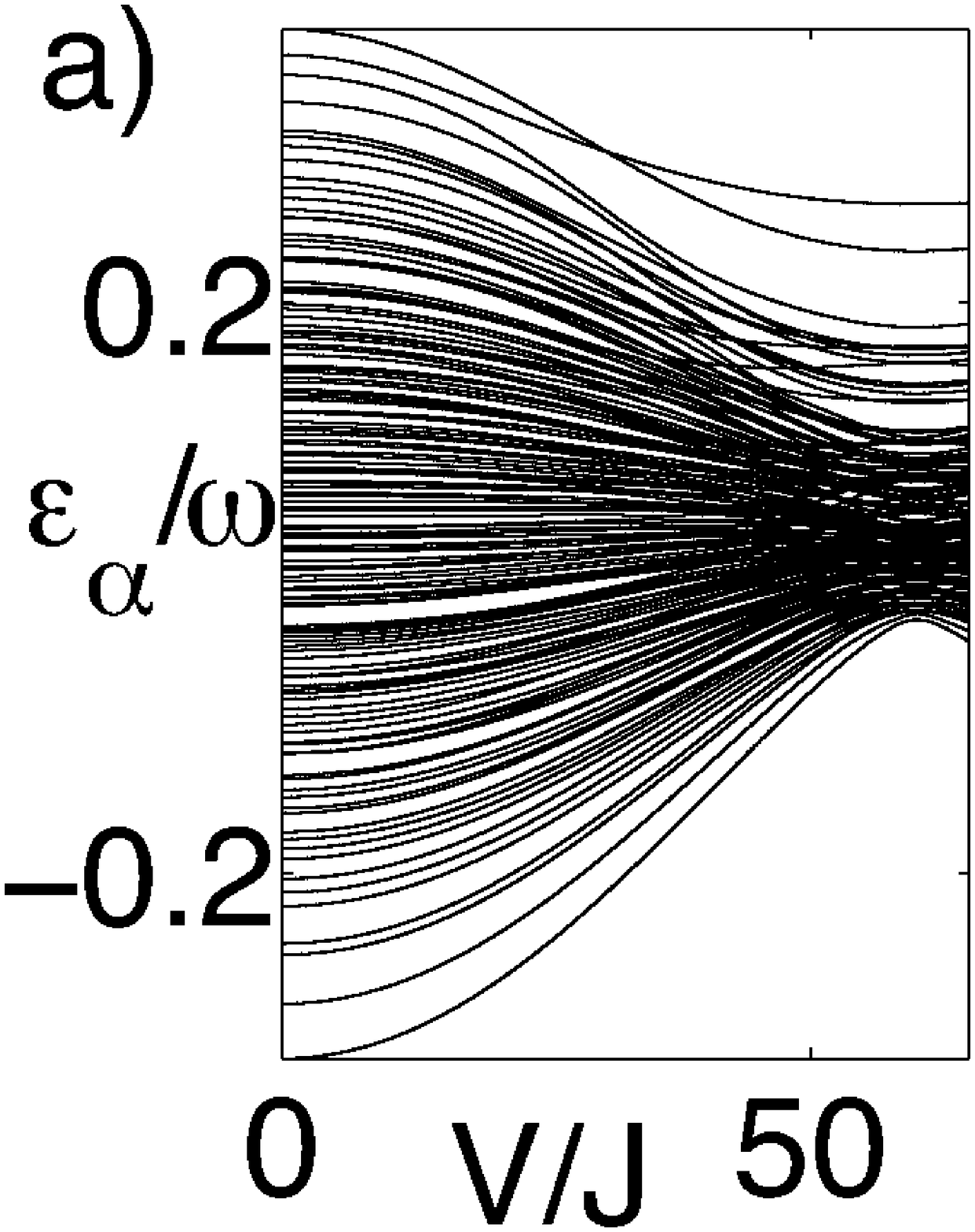,width=0.35\columnwidth} & \epsfig{file= 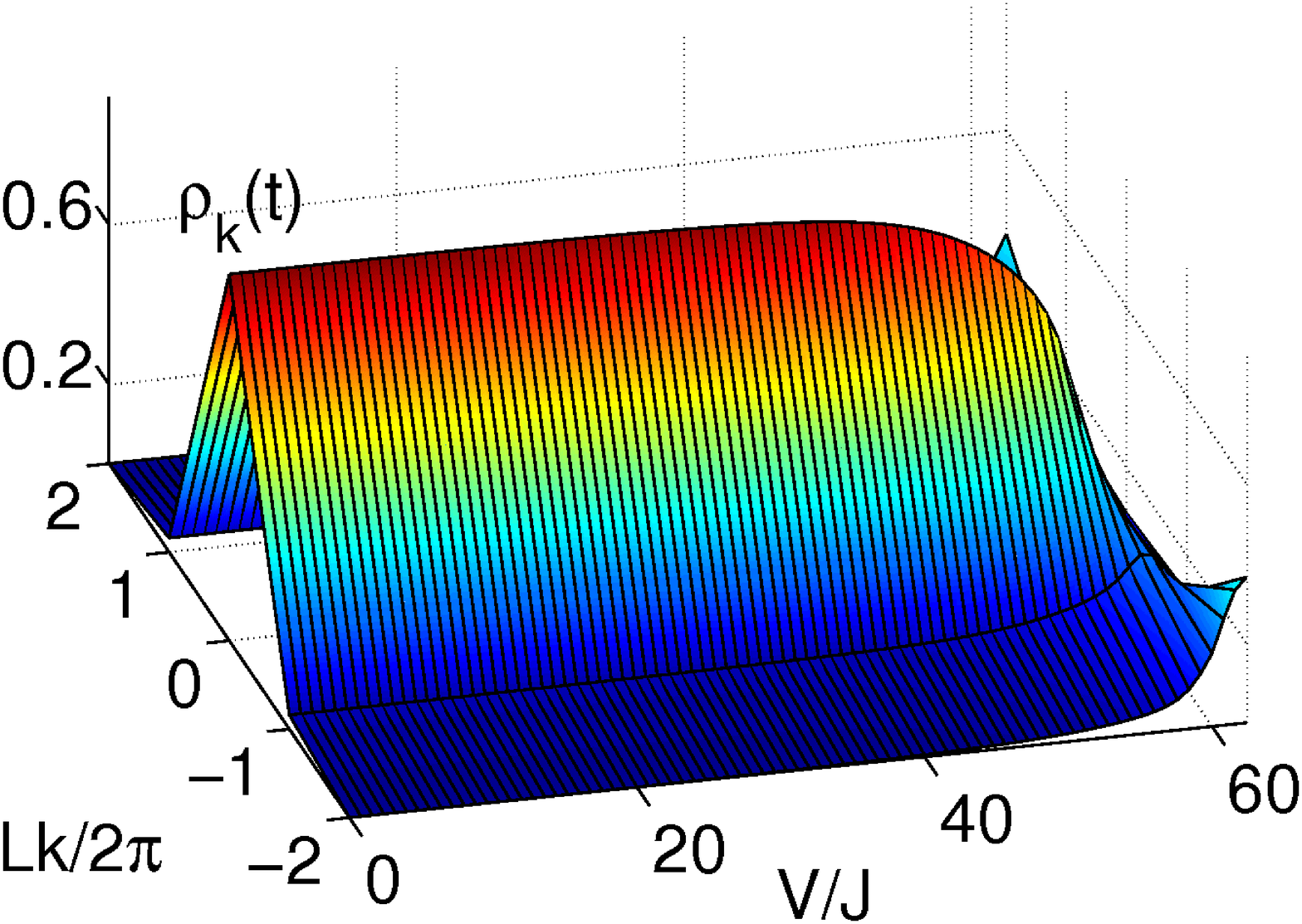,width=0.65\columnwidth} 
\end{tabular}
\caption{(Color online) a) Quasienergy levels for $N=L=5$, $U/J=0.5$, $\omega/J=25$, $W/J=1$. b) Evolution of the momentum distribution of the ground state for a potential with a linear ramp of step $\Delta V/J=0.01$. }
\label{fig:sfbg}
\end{figure}

\subsection{Intermediate disorder $W/U \sim 1$}
For stronger values of disorder, $W/U \sim 1$, the predicted ground states of the disordered BH model for $U/J\gg 0$ are BG states \cite{Burnett03,ReviewMaciej}. These states are insulating, with a spatial distribution which follows the disorder potential pattern, and show a continuous spectrum.

\begin{figure}
\centering
\begin{tabular}{cc}
\epsfig{file= 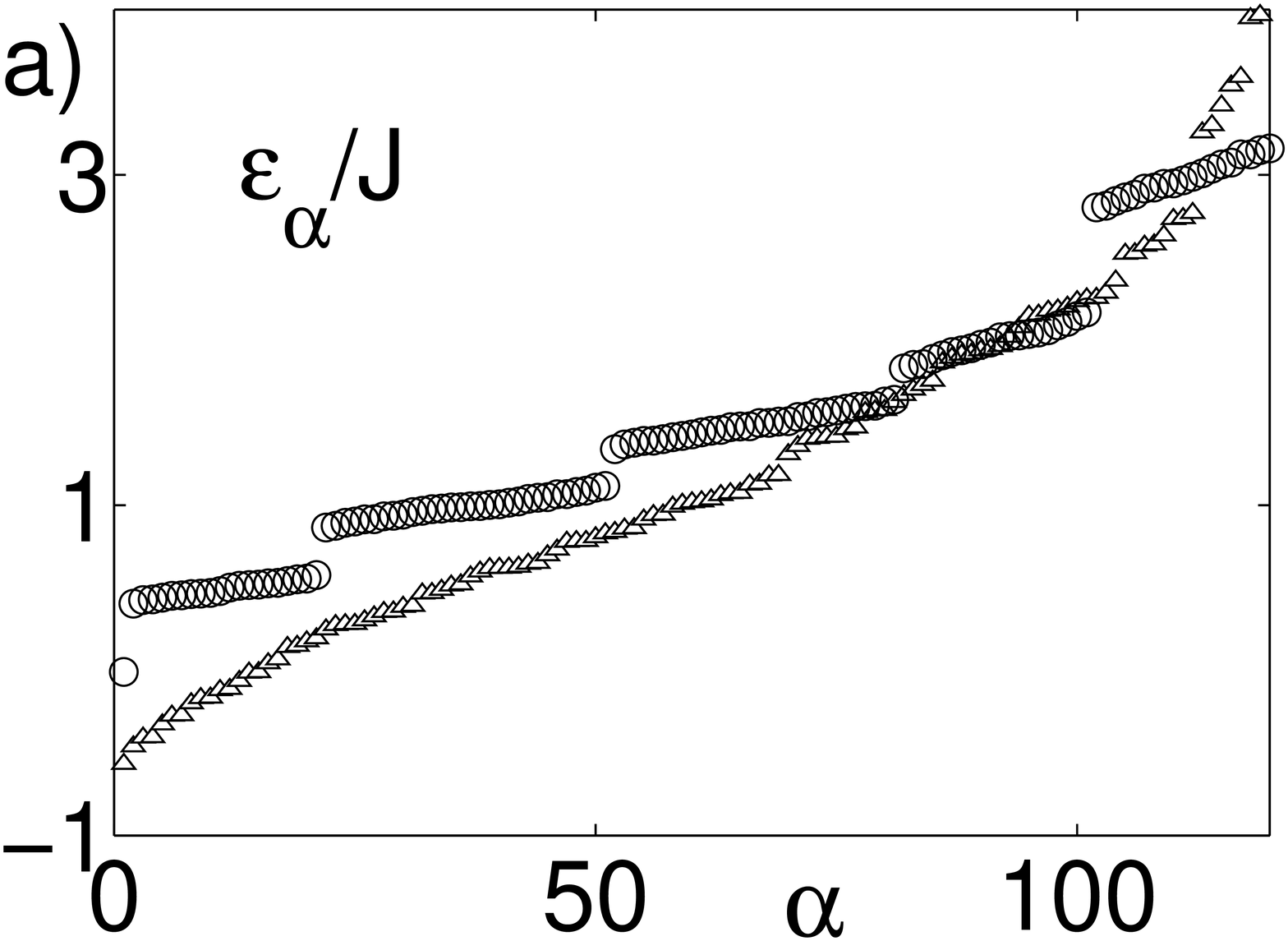,width=0.5\columnwidth} &  \epsfig{file= 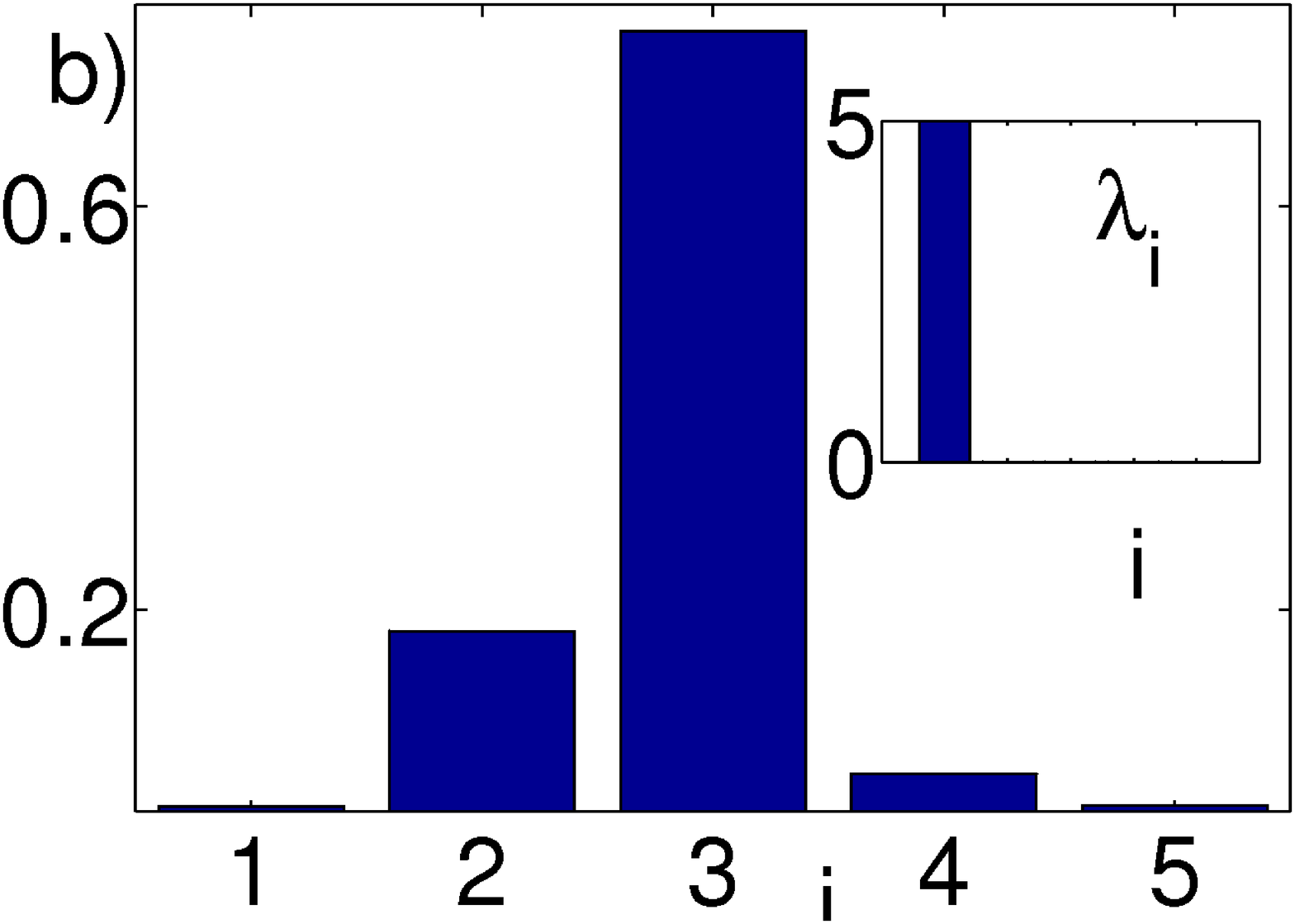,width=0.5\columnwidth} 
\end{tabular}
\caption{(Color online) a) Quasienergy spectrum $\varepsilon_\alpha$ ($\alpha$ indicates the quasienergy ordering) of $H(t)$ when $V/\omega=2.4$ such that $J_{eff}=0$ for $W/J=1$ ($\bigtriangleup$) and $W/J=0.1$ ($\bigcirc$). The continuous line for $W/U=2$ resembles a BG-like behavior, while the discrete spectrum obtained for $W/U=0.2$ resembles a MI-like spectrum. ($N=L=5$ and $U/J=0.5$) b) Inset: Eigenvalues of the one-particle density matrix of the ground state of the initial Hamiltonian at $t=0$ ($H_0$) with $U/J=0.05$,$W/J=7$ shown in fig. 3 e). Square of the most populated eigenfunction of the one-particle density matrix in fig. 3 e).}
\label{fig:spec}
\end{figure}

Our time-dependent calculations show that we can drive an adiabatic transition from an initial SF-like state, the ground state of $H$ at $t=0$, into a BG-like state at a final time when the effective tunneling $J_{\rm eff}\sim 0$. The quasienergy spectrum and the momentum distribution of the evolved state are shown in figure \ref{fig:sfbg}. The one-particle density matrix smoothly changes from a flat structure peaked around the deepest well induced by disorder (see fig. \ref{fig:dens} c)) into a diagonal shape for $V/\omega$ close to the first zero of the Bessel function Eq. (\ref{eq:Jbessel}) as shown in fig. \ref{fig:dens} d). To completely characterize the BG-like state obtained in this manner from the MI-like state obtained in the previous section for smaller values of the disorder we show in fig. \ref{fig:spec} the spectrum obtained when the tunneling is coherently suppressed for $V/\omega=2.4$. We observe the gapped MI spectrum for $W/U=0.2$ and the continuous spectrum for the BG obtained with $W/U=2$. 

We now look at the effect of the occupation number on the adiabatic time scale. 
For commensurate filling, the ground state which was always gapped from the excited states for small values of disorder, becomes now closer to the higher excited states and the time scale needed for adiabatic ramping increases. For the particular example we have calculated ($N=5$ shown in fig. \ref{fig:sfbg}), we need to go to larger time scales $\Delta V/J=0.01$ for $W/U=2$ in order to observe adiabatic behavior. Note however in fig. \ref{fig:schemes} c) that the lowest energy state is still gapped across the transition. For non-commensurate filling, disorder breaks the degeneracy of all the Fock states with similar occupation patterns for the effective Hamiltonian with $J_{\rm eff}=0$. Increasing disorder strength implies increasing gap and we thus obtain faster ramping scales. For $N=4$ we obtain $\Delta V/J=0.01$ for $W/U=2$, one order of magnitude faster than for $W/U=0.2$ and of the same order than in a system with $N=5$ and $W/U=2$. This is in contrast to the small disorder case in Sec.\ref{sec:weak}, where the ramping depends strongly on the filling factor.

We expect this behavior to persist even for larger system sizes because in a disordered potential the sites with minimum energy, $\sim -W/2$, are randomly distributed across the lattice. The first excited states correspond to particles distributed in those minimum energy sites which in principle are far away in the lattice and cannot be coupled to low orders by the nearest neighbor tunneling term.

\begin{figure}
\centering
\begin{tabular}{cc}
\epsfig{file= 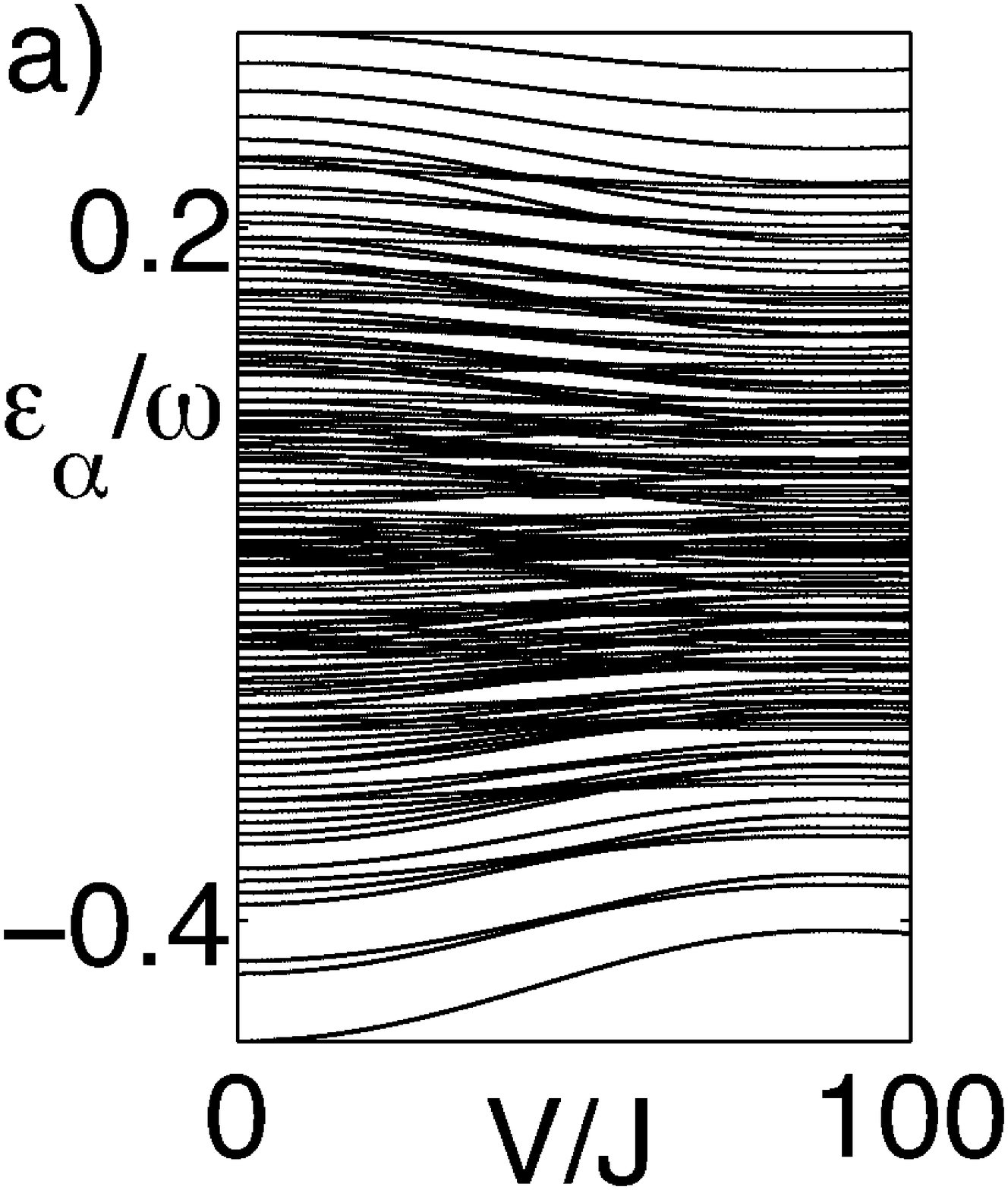,width=0.35\columnwidth,clip=} &
\epsfig{file= 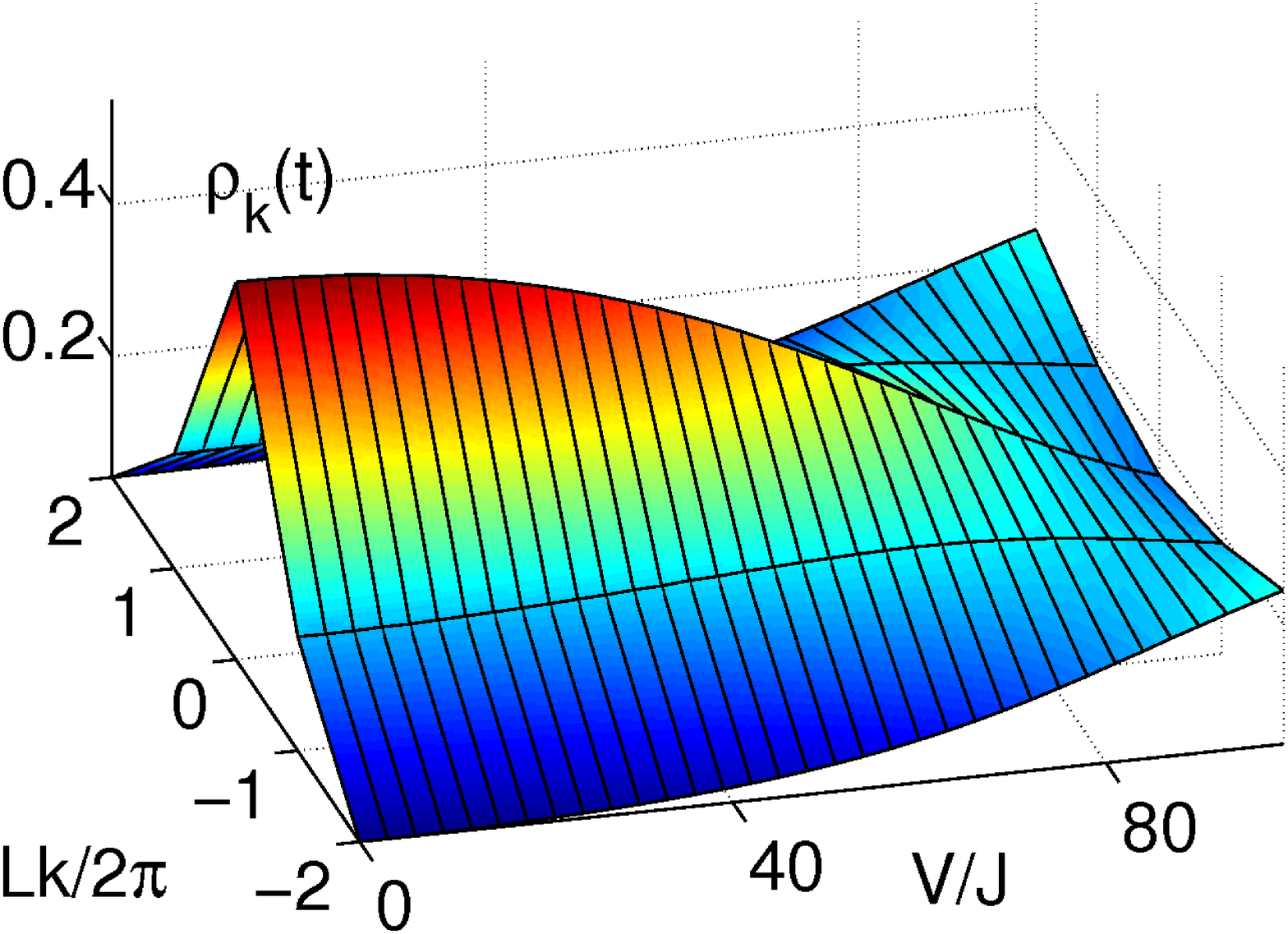,width=0.65\columnwidth,clip=} 
\end{tabular}
\caption{(Color online) a) Quasienergy level structure for $N=L=5$, $U/J=0.05$, $\omega/J=37$, $W/J=7$. b) Evolution of the momentum distribution of the ground state for a potential with a linear ramp of step $\Delta V/J=0.3$. }
\label{fig:agbg}
\end{figure}
\subsection{Strong disorder $W/U \gg 1$}

In the regime where the disorder dominates, the condensate state which appears for large values of $J/U$ is not extended over the whole space, contrary to previous cases. For $J/U$ large, the ground state is a condensate localized in the region of the lattice most distorted by the disorder. This localized condensed state which appears in our small system resembles the properties of AG states appearing in larger systems \cite{Burnett03}, where particles condense into localized Anderson states with an exponential decay distributed across the lattice in different islands. We calculate the one-particle density matrix of the initial ground state for $W/U=140$ shown in fig. \ref{fig:dens} e). The eigenvalues of the one-particle density matrix (inset of fig.\ref{fig:spec} b)) show that only one state is macroscopically occupied. This state, shown in fig.\ref{fig:spec} b), decays exponentially in space. The quasienergy spectrum and the momentum distribution of the evolved state depicted in fig. \ref{fig:agbg}, show that the system evolves from a state with finite condensate fraction into an insulating state. Note that for bigger system sizes, both the initial and final states are insulating, and our fast oscillatory force is able to drive a transition from states with off-diagonal terms in the one-particle density matrix around the deepest sites of the lattice into states with only diagonal terms in the one-particle density matrix, as shown in fig.\ref{fig:dens}. This final diagonal states appearing for $J_{\rm eff}\sim 0$ resemble BG insulating states.

The one-particle density matrix shown in fig. \ref{fig:dens} peaks around the deepest lattice site and looses its off-diagonal elements when the system becomes maximally localized. For $N=5$ we observe that the ground state remains always the ground cyclic state of the quasienergy structure shown in fig.\ref{fig:sfbg} for ramps with $\Delta V/J=0.3$ which would mean $t_{\rm ramp}\sim 10^{-2}$s. In this regime when $W/J$ is large, the quasienergy spectrum in fig. \ref{fig:agbg} is dominated by disorder for all values of the ramping potential $V/J$. Even at $t=0$ when the tunneling is largest $J_{\rm eff}=J$ the hopping is mainly taking place around the deepest potential well.

For larger system sizes, where the deepest sites are in principle far away, the ground state would be a proliferation of islands of the type considered here. The near energy states appearing in the spectrum as the system size is increased are not connected by the nearest neighbor tunneling matrix term of the Hamiltonian and one can therefore expect that the AG-BG-type transition on each island takes place independently.  We thus expect that the transition between ground states can be realized adiabatically within experimentally feasible time scales. 

\subsection{System size considerations}
When the system size increases the quasienergy spectrum becomes more and more dense. The requirement of high frequencies is that $\omega \gg U,J,W$ and thus when $N$ increases the quasienergies of the high energy states can cross that of the ground state; i.e. the frequency $\omega$ may in principle induce resonances between different states. Yet the states which are coupled by $\omega$ have very different occupation number patterns and do not have any effective overlap. This type of accidental crossings are traversed diabatically for any physically relevant time scale as discussed in \cite{Holthaus06}.

All the adiabatic time scales calculated in this work involve very small particle numbers. Due to computational restrictions, any exact time-dependent calculation for the BH model involves small number of particles. This limitation has not prevented small calculation predictions to be proven in real experiments with much larger number of particles \cite{Holthaus06,Arimondo09} and no disorder. The disorder potential we consider is white noise of strength $W$ which means that for a large system the average energy gap between neighboring sites (the ones coupled by the Hamiltonian) is of order $W/2$. In other words, for a large system, lattice sites distorted by $-W/2$ are very far on average. Thus, most energy levels which appear close to the ground state cannot be coupled to it to low orders indicating that our small system calculations can be extrapolated to larger system sizes.

\section{Conclusions\label{sec:concl}}

We have shown that the adiabatic ramping of a constant force rapidly oscillating in time allows for the complete scan of the parametric dependence of the ground state of a {\it disordered} BH model. As shown previously \cite{Holthaus06}, the inclusion of an ac field in the Bose-Hubbard Hamiltonian is equivalent to a renormalization of the effective tunneling for the time independent spatial Hamiltonian. A succession of oscillatory fields where the amplitude is slowly increased gives rise in the adiabatic limit to a succession of effective Hamiltonians from some initial $J$ to the final $J_{\rm {eff}}=0$ and thus allows for a complete scan of the parameters $U/J$ and $W/J$ with $W/U$ constant. 

We have found that in spite of the increasing complexity of the quasienergy diagram with disorder, adiabatic following of the ground state is feasible with a similar time scale as in a system without disorder. Adiabatic following of the ground state enables the transition from states with long-range coherence into localized states both for commensurate and non-commensurate fillings of the lattice potential.  For strong disorder we have shown that the transition between insulating states localized in the most distorted regions of the lattice is possible. 
We have found that the addition of disorder shortens the time scale needed for adiabatic ramping if the disorder is weak or strong compared to the interaction strength.  When the disorder is comparable to the interaction strength, adiabatic behavior is still feasible. 

Our results indicate that this type of transitions between different regions of the phase diagram can be realized adiabatically for experimentally feasible time scales in systems with large number of particles. The adiabatic ramping of a fast oscillatory force could be used not only to control localization of disordered bosons but also to scan richer phase diagrams predicted for bosonic or fermionic systems in optical lattices.

%acknowledgements
This work was supported by the Spanish Government through projects FIS 2007-616866 and FIS2006-12783-C03-01. M R acknowledges support through the Ramon y Cajal programme. RAM's contract is financed through the I3P by CSIC and the European Commission.


\begin{thebibliography}{cc}
\bibitem{SFMI}
D. Jaksch {\it et al.}, Phys.\ Rev.\ Lett.\ \textbf{81}, 3108
(1998); M. Greiner {\it et al.}, Nature (London) \textbf{415}, 39 (2002).
\bibitem{Holthaus06}
A. Eckardt, C.  Weiss, and M. Holthaus, Phys. Rev. Lett. \textbf{95}, 260404 (2005).


\bibitem{Arimondo09}
A. Zenesini, H. Lignier, D. Ciampini, O. Morsch and E. Arimondo, Phys.\ Rev.\ Lett.\ \textbf{102}, 100403 (2009).

\bibitem{HolHone96} M. Holthaus and D. W.
Hone, Phil. Mag. B {\bf 74}, 105 (1996).
\bibitem{Korsch}
T. Hartman, F. Keck,H. J. Korsch and S. Mossmann, New J. Phys. \textbf{6}, 2 (2004); B. M. Breid, D. Witthaut and H. J. Korsch, New J. Phys. \textbf{8}, 110 (2006).
\bibitem{Dunlap88} D. H. Dunlap and V. M. Kenkre 
Phys. Rev. B {\bf 34}, 3625 (1986).


\bibitem{Holthaus95} M. Holthaus, G. H. Ristow, and D.W. Hone,
Phys. Rev. Lett. {\bf 75}, 3914 (1995).




\bibitem{DeMarco09}
M. White, M. Pasienski, D. McKay, S. Q. Zhou, D. Ceperley and B. DeMarco, Phys.\ Rev.\ Lett.\ \textbf{102}, 055301 (2009).
\bibitem{Inguscio}
L. Fallani, J. E. Lye, V. Guarrera, C. Fort and M. Inguscio, Phys. Rev. Lett. \textbf{98}, 130404 (2007).
\bibitem{Fisher89}
M. P. A. Fisher, P. B. Weichman, G. Grinstein and D. S. Fisher, Phys.\ Rev.\ B, \textbf{40}, 546 (1989).
\bibitem{Damski03}
B. Damski, J. Zakrzewski, L. Santos, P. Zoller and M. Lewenstein, Phys.\ Rev.\ Lett.\ \textbf{91}, 080403 (2003).
\bibitem{Burnett03}
R. Roth and K. Burnett, Phys. Rev. A \textbf{68},023604 (2003); R. Roth and K. Burnett, J. Opt. B \textbf{5}, S50 (2003).
\bibitem{Kruger09}
F. Kr\"uger, J. Wu, and P. Phillips, Phys. Rev. B. \textbf{80}, 094526 (2009).
\bibitem{ReviewMaciej}
M. Lewenstein, A. Sanpera, V. Ahufinger, B. Damski, A. Sen (De), U. Sen
Adv. Phys. \textbf{56}, 243 (2007).      

\bibitem{Raizen98}  K. W. Madison, M. C. Fischer, R. B. Diener, Qian Niu, and M. G. Raizen
Phys. Rev. Lett. \textbf{81}, 5093 (1998).

\bibitem{HolthausEPL}
A. Eckardt and M. Holthaus, Eur. Phys. Lett. \textbf{80}, 50004 (2007).

\bibitem{Anderson}
P. W. Anderson, Phys. Rev. \textbf{109}, 1492 (1958).

\bibitem{Sambe} H. Sambe, Phys. Rev. A {\bf 7}, 2203 (1972).

\bibitem{Dittrich91} F. Grossmann, T. Dittrich, P. Jung, P.
H\"anggi, Phys. Rev. Lett. \textbf{67}, 516 (1991).

\bibitem{Holthaus92} M. Holthaus, Phys. Rev. Lett. {\bf 69}, 351 (1992).
\bibitem{Creffield08}
C. E. Creffield and F. Sols, Phys. Rev. Lett. \textbf{100}, 250402 (2008).
\bibitem{Molina06a} D.F. Martinez, R.A. Molina, Phys. Rev. B {\bf 73}, 073104 (2006). 
\bibitem{Molina06b} D.F. Martinez, R.A. Molina, Eur. Phys. J. B {\bf 52}, 281 (2006).
\bibitem{Lanczos}
T. J. Park, J. C. Light, J. Chem. Phys. \textbf{85}, 5870 (1986).

\end{thebibliography}
\end{document}